\documentclass[useAMS,usenatbib]{mn2e}

\usepackage{psfig}
\include{epsf}
\newcommand{\be}{\begin{equation}}
\newcommand{\ee}{\end{equation}}
\newcommand{\ba}{\begin{eqnarray}}
\newcommand{\ea}{\end{eqnarray}}
\newcommand{\nn}{\nonumber}
\def\simless{\mathbin{\lower 3pt\hbox
   {$\rlap{\raise 4pt\hbox{$\char'074$}}\mathchar"7218$}}}
\def\simgreat{\mathbin{\lower 3pt\hbox
   {$\rlap{\raise 4pt\hbox{$\char'076$}}\mathchar"7218$}}}   

\title[The UKIDSS EDR]{The UKIRT Infrared Deep Sky Survey Early Data
Release} \author[S. Dye et al.]  {S. Dye$^1$\thanks{E-mail:
s.dye@astro.cf.ac.uk}, 
S. J. Warren$^2$, 
 N. C. Hambly$^3$, 
N. J. G. Cross$^3$,
S. T. Hodgkin$^4$, \newauthor
M. J. Irwin$^4$,  
A. Lawrence$^3$, 
A. J. Adamson$^5$, 
O. Almaini$^6$, 
A. C. Edge$^7$,\newauthor
P. Hirst$^5$, 
R. F. Jameson$^8$, 
P. W. Lucas$^9$, 
C. van Breukelen$^{10}$, 
J. Bryant$^3$, \newauthor
M. Casali$^{11}$,
R. S. Collins$^3$, 
G. B. Dalton$^{10}$, 
J. I. Davies$^1$,
C. J. Davis$^5$,  \newauthor
J. P. Emerson$^{12}$,
D. W. Evans$^4$,
S. Foucaud$^6$, 
E. A. Gonzales-Solares$^4$, \newauthor
P. C. Hewett$^4$, 
T. R. Kendall$^9$, 
T. H. Kerr$^5$,
S. K. Leggett$^5$, 
N. Lodieu$^8$,   \newauthor
J. Loveday$^{13}$, 
J. R. Lewis$^4$,
R. G. Mann$^3$,
R. G. McMahon$^4$, 
D. J. Mortlock$^2$, \newauthor
Y. Nakajima$^{14}$, 
D. J. Pinfield$^9$,   
M. G. Rawlings$^5$, 
M. A. Read$^3$,  
M. Riello$^4$,  \newauthor 
K. Sekiguchi$^{15}$, 
A. J. Smith$^{13}$, 
E. T. W. Sutorius$^3$, 
W. Varricatt$^5$, 
N. A. Walton$^4$, \newauthor 
S. J. Weatherley$^{16}$ 
\vspace{7mm}\\
$^1$Cardiff University, School of Physics \& Astronomy, Queens Buildings,
The Parade, Cardiff, CF24 3AA, U.K. \\
$^2$Astrophysics Group, Imperial College London, Blackett Laboratory,
Prince Consort Road, London, SW7 2AZ, U.K.\\
$^3$Royal Observatory Edinburgh, Blackford Hill, Edinburgh, EH9 3HJ,
U.K.\\
$^4$Institute of Astronomy, Madingley Rd., Cambridge, CB3 0HA, U.K.\\
$^5$Joint Astronomy Centre, 660 N. A'ohoku Place, University Park,
Hilo, Hawaii 96720, U.S.A.\\
$^6$School of Physics and Astronomy, University of Nottingham,
University Park, Nottingham, NG7 2RD, U.K. \\ 
$^7$Department of Physics, Durham University, South Road, DH1 3LE,
U.K. \\
$^8$Department of Physics and Astronomy, University of Leicester,
Leicester, LE1 7RH, U.K. \\
$^9$Centre for Astrophysics Research, Science and Technology Research 
Institute, University of Hertfordshire, Hatfield, AL10 9AB, U.K. \\
$^{10}$Department of Physics,  Denys Wilkinson Building,
Keble Road, Oxford, OX1 3RH, U.K.  \\
$^{11}$ESO, Karl-Schwarzschild-Str. 2, D-85748 Garching bei M\"{u}nchen, 
Germany \\
$^{12}$Astronomy Unit, School of Mathematical Sciences, Queen Mary, 
University of London, Mile End Road, London E1 4NS \\
$^{13}$Department of Physics and Astronomy, University of Sussex,
Brighton, East Sussex,  BN1 9QH, U.K. \\
$^{14}$National Astronomical Observatory of Japan,
2-21-1 Osawa, Mitaka, Tokyo, 181-8588, JAPAN \\
$^{15}$Subaru Telescope, National Astronomical Observatory of Japan,
650 North A'ohoku Place, Hilo, HI 96720, U.S.A. \\
$^{16}$Detica, Surrey Research Park, Guildford, Surrey, GU2 7YP, U.K.
}

\begin{document}

\date{Document in prep.}

\pagerange{\pageref{firstpage}--\pageref{lastpage}} \pubyear{2006}

\maketitle

\label{firstpage}

\begin{abstract}
This paper defines the UKIRT Infrared Deep Sky Survey (UKIDSS) Early
Data Release (EDR). UKIDSS is a set of five large near-infra-red
surveys defined by \citet{lawrence06}, being undertaken with the UK
Infra-red Telescope (UKIRT) Wide Field Camera (WFCAM). The programme
began in May 2005 and has an expected duration of seven years. Each
survey uses some or all of the broadband filter complement
$ZYJHK$. The EDR is the first public release of data to the European
Southern Observatory (ESO) community.  All worldwide releases occur
after a delay of 18 months from the ESO release. The EDR provides a
small sample dataset, $\sim 50$ deg$^2$ (about $1\%$ of the whole of
UKIDSS), that is a lower limit to the expected quality of future
survey data releases. In addition, an EDR+ dataset contains all EDR
data plus extra data of similar quality, but for areas not observed in
all of the required filters (amounting to $\sim 220$ deg$^2$). The
first large data release, DR1, will occur in mid-2006.  We provide
details of the observational implementation, the data reduction, the
astrometric and photometric calibration, and the quality control
procedures. We summarise the data coverage and quality (seeing,
ellipticity, photometricity, depth) for each survey and give a brief
guide to accessing the images and catalogues from the WFCAM Science
Archive.
\end{abstract}


\section{Introduction}

UKIDSS is the UKIRT Infrared Deep Sky Survey \citep{lawrence06},
carried out using the Wide Field Camera \citep[WFCAM;][]{casali06}
installed on the United Kingdom Infrared Telescope (UKIRT). Survey data
acquisition started in May 2005. This paper defines
the UKIDSS Early Data Release (EDR), the first release of UKIDSS data
products to the ESO-wide astronomical community. The data were
released on 10th February 2006, and are available from {\tt
http://surveys.roe.ac.uk/wsa}.

UKIDSS is a programme of five individual surveys that each use some or
all of the broadband filter complement $ZYJHK$ \citep[described
by][]{hewett06}, and that span a range of areas and depths. There are
three high galactic latitude surveys; the Large Area Survey (LAS), the
Deep Extra-galactic Survey (DXS), and the Ultra Deep Survey (UDS),
covering complementary combinations of area and depth. Besides the
LAS, there are two other wide surveys at low galactic latitudes, aimed
at targets in the Milky Way; the Galactic Plane Survey (GPS), and the
Galactic Clusters Survey (GCS). The three wide, shallow surveys, LAS,
GCS and GPS, use a total integration time of 40s or 80s per field per
filter. The layout and filter complement of each survey are provided
in Section
\ref{sec_2year_plan}. The complete UKIDSS programme is scheduled to
take 7 years, requiring $\sim$1000 nights on UKIRT. The course of the
full 7 year programme will be guided, to a certain extent, by the
progress and scientific findings of the first two years of
surveying. The design of these first two years consists of a set of
fixed, shorter-term goals that define the `2--year plan'.

The purpose of the present paper is to provide a self--contained guide
to the EDR. We have therefore included relevant background
information extracted from the set of UKIDSS reference technical
papers on; the surveys \citep{lawrence06}, the photometric system
\citep{hewett06}, the camera \citep{casali06}, the data pipeline
\citep{irwin06}, and the data archive \citep{hambly06},
compressing details that will be the same for all data releases, and
expanding on details that are specific to the EDR. The name EDR is
copied from the Sloan Digital Sky Survey (SDSS) EDR
\citep{stoughton02} since the UKIDSS EDR serves a similar purpose to
the SDSS EDR, i.e. it is a prototype of future larger
releases. Similarly, the introduction to UKIDSS given by
\citet{lawrence06}  mirrors the purpose of 
\citet{york00}, introducing SDSS. We will also follow SDSS by
naming the first large data release `DR1'. The SDSS and 2 Micron All Sky
Survey \citep[2MASS;][]{skrutskie06} have both fundamentally influenced the
design of all elements of UKIDSS.

The outline of the paper is as follows: In the next section, we
discuss the background to the UKIDSS programme.  In Section
\ref{sec_wfcam}, we provide a brief description of WFCAM, and in
Section \ref{sec_2year_plan} we give an overview of the UKIDSS 2-year
goals, in terms of area and filter coverage for each survey.  We then
follow the data-train sequence. In Section \ref{sec_implement} we
provide details of the observational implementation of the surveys, at
a level that is required for understanding the data. For example, this
includes offsetting, microstepping and stacking (i.e. averaging)
strategies which dictate the structure of the image files. Section
\ref {sec_pipeline} describes the data reduction pipeline, and the
procedures followed for the astrometric and photometric calibration of
the data. Section \ref {sec_data_flaws} catalogues a variety of
artifacts that can occur in the data. The output from the pipeline is
a set of reduced, stacked frames and associated catalogues of detected
objects. After ingestion to the archive, an initial series of quality
control (QC) tests reject sub-standard data before the catalogues
across different bands are merged. A second round of QC checks are
then applied to the merged catalogues.  Section \ref{sec_qc} describes
these QC procedures. The contents of the EDR are described in Section
\ref{sec_edr}, where, in particular, plots of the distribution of the
fields on the sky are provided. Access to the EDR is through the WFCAM
Science Archive (WSA), which is described in Section \ref{sec_wsa}.

\section{UKIDSS Background}

The design and implementation of the UKIDSS programme is the
responsibility of the UKIDSS Consortium, an assemblage of astronomers
with a common aim of completing the surveys.  Observing is shared
between UKIRT staff and UKIDSS astronomers. UKIDSS is an ESO public
survey and UKIDSS consortium members have no proprietary data
rights. Acquired data are shipped from UKIRT to the Cambridge
Astronomical Survey Unit (CASU) at weekly intervals. At this stage, a
copy of the raw UKIDSS data is sent from CASU to the ESO archive. At
CASU, the data are processed by the WFCAM pipeline to produce stacked
image data and source catalogues (see Section \ref{sec_pipeline}). The
processed data products are electronically transferred to the Wide
Field Astronomical Unit (WFAU) in Edinburgh where they are ingested by
the WSA (see Section \ref{sec_wsa}). All WFCAM data, whether UKIDSS or
not, follows the same route. UKIDSS, being the largest WFCAM user, has
interacted closely on the design and development of the pipeline and
archive.

Following commissioning of WFCAM, a few nights (approximately 10 hours
per survey) were devoted to science verification (SV) observations, in
order to test the entire data train made up of the three elements: 1)
the observational strategies employed by the different surveys (e.g.
to check that integration times give background limited observations,
offset patterns enable good sky correction and guide stars are
sufficiently bright), 2) the processing steps executed by the WFCAM
pipeline, from flatfielding and sky correction of the image data, to
stacking and catalogue generation, 3) the functionality of the WSA,
both archive ingest and source merging, and retrieval of the image and
catalogue data through the web-based user interface (see Section
\ref{sec_wsa}). The SV data were analysed by UKIDSS consortium
members, the results were fed back to CASU and WFAU, and the process
was repeated after adjustments to the pipeline and archive. On the
basis of the SV results, a set of QC procedures for
UKIDSS data was agreed (described in Section \ref{sec_qc}) and
then applied to the EDR by WFAU. The SV dataset, much smaller
than the EDR, is mostly heterogeneous. With the exception of some DXS
data, the SV data have not been included in the EDR.

WFCAM is scheduled in blocks, a few months at a time, separated by
intervals of a few months, and in such a way as to ensure that over a
two-year span, access over the entire RA range is reasonably
uniform. UKIDSS uses about $40\%$ of all UKIRT time. The first WFCAM
survey block, hereafter 05A, included UKIDSS survey observations over
several nights commencing 13th May and ending 19th June 2005. All data
from 05A that passes QC is released here. Additionally some DXS data
taken in the science verification phase, back to 9th April have been
included. The second block, hereafter 05B, commenced 27th Aug 2005 and
ended 25th January 2006. DXS and UDS data taken in 05B up to 27th
September 2005, that satisfy QC, are also included in this
release. Some changes to observing procedures, detailed in Section
\ref{sec_implement}, were made between semesters 05A and 05B. Further
changes are not anticipated.

The 05A data are somewhat inferior to the 05B data in terms of image
quality and cosmetics. Correct alignment of the instrument requires
that the instrument focal plane assembly and optical focal plane be
co-planar to within a small fraction of a milliradian. This can only
be measured on sky, but requires disassembly of the instrument to
adjust properly. An interim adjustment can be made by tilting the
secondary mirror of the telescope as this results in a corresponding
tilt of the optical focal plane, though this introduces other
aberrations into the optics. The 05A data were taken with this type of
interim adjustment in place and in good seeing it is evident that the
data are compromised by this. The instrument was opened up and the
alignment corrected internally between the 05A and 05B observing
blocks.  Secondly, there is an issue of out--of--focus images of dirt
on the field lens (see Section \ref{sec_data_flaws}), which forms the
end of the barrel of the instrument. Because the lens lies close to an
intermediate focus, out--of--focus images of dust or marks on the
lens are visible in some frames. The top surface is frequently
cleaned, while the bottom surface was cleaned in the interval between
the 05A and 05B blocks to remove marks left by a lens handling jig,
visible in the 05A data.

The EDR dataset is relatively large compared to existing near-infra-red
surveys (about as many photons have been collected as in 2MASS), and
will be useful for science exploitation. However, compared to the
whole of UKIDSS the EDR is only a small fraction, about $1\%$. From
the UKIDSS perspective, the EDR is therefore a small prototype sample,
released to the community as a stepping stone toward the goal of
routine and prompt release of survey-quality data. We welcome feedback
on the quality of the data and the functionality of the WSA. We aim to
finalise the details of the data-train that define survey-quality
products over early 2006, with further refinements of the pipeline,
as well as adjustments and additions to the QC procedures. The entire
05A and 05B data will then be prepared for the first large data
release, DR1, scheduled for  mid-2006. With this in mind, and
considering the image quality and cosmetic issues of the 05A data, we
have relaxed the QC criteria to a limited extent (e.g. depth and
seeing), with the consequence that some fields included in the EDR
will be excluded from DR1 and instead will be re-observed.

The EDR appears in the WSA as two databases. The EDR database is all
the data where, for any pointing, the images and catalogues in all
filters for that survey are available and pass QC. The EDR+ database
is all of the EDR data plus the remainder that passes QC, i.e.
pointings where the full complement of filters for that survey is
incomplete, either because some remain to be observed, or because some
failed QC. In the remainder of the paper, where it is appropriate to
make the distinction, we refer to the `EDR database' and the `EDR+
database'. The general term `EDR' implies the combination.  The extra
content of the EDR+ database is very heterogeneous in terms of
completeness of the filter complement so that its usefulness is less
apparent, considering that the surveys were designed around science
goals that require the complete filter complement. We therefore
anticipate that users will be most interested in the EDR database.
Accessing the two databases is explained in detail in Section \ref{sec_wsa}.

The first UKIDSS observations were made in May 2005, so that EDR
represents a maximum lag of 9 months from acquisition to release.
After finalisation of the pipeline and QC procedures, and automation
of the latter, the time-lag will rapidly diminish. 
At present the policy is that a release will occur once all data from
an observing block have been processed. Ideally the delay from
completion of the observing block to release of the data will become
as short as 3 months. Each release will be accompanied by an
announcement publication defining the specific data products included
in that release.

\section{WFCAM}
\label{sec_wfcam}

WFCAM was built by the UK Astronomical Technology Centre (UK ATC) in
Edinburgh as a common user instrument for UKIRT. The UKIDSS programme
owes its feasibility to the large field of view of the camera. This
section provides details of the design and construction of WFCAM,
relevant for understanding the contents of the EDR, including some of
the unusual properties of WFCAM data, as well as the characteristics
peculiar to this release. We refer the reader to \citet{casali06} for
full technical details of WFCAM.

\begin{figure*}
\epsfxsize=100mm
{\hfill
\epsfbox{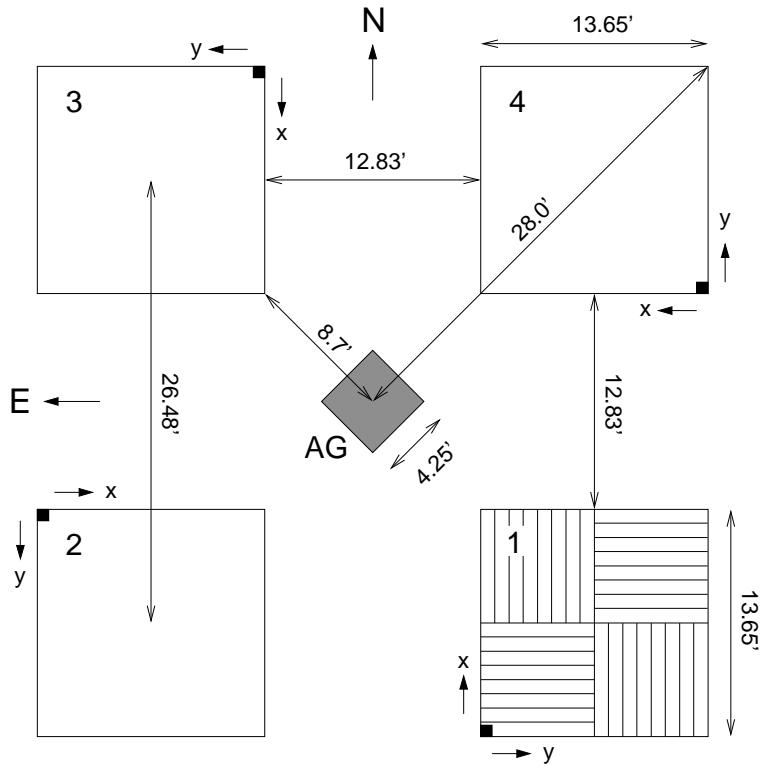}
\hfill}
\caption{Layout of the WFCAM focal plane. Detector numbering follows
that used in the WSA and is the same as the extension numbering in the
FITS data files. Detectors are divided into four quadrants, each
quadrant having 8 channels as illustrated in detector 1. The autoguider 
(grey shading) sits at the centre of the focal plane. North is
up, East is left. }
\label{wfcam_focal_plane}
\end{figure*}


The survey speed of an instrument is proportional to the
telescope+instrument \'{e}tendue\footnote{product of telescope
collecting area, and solid angle of instrument field of view,
sometimes called {\em grasp}}. At the time of writing, WFCAM's
\'{e}tendue of $2.38\,$m$^2\,$deg$^2$ is the largest of any
near-infrared imager in the world. The large field of view of WFCAM,
0.21 deg$^2$, is achieved with four Rockwell Hawaii-II $2048 \times
2048$ $18\mu$m-pixel array detectors, with a relatively large pixel
scale of $0.4\arcsec$. In good seeing the point spread function (PSF)
is consequently under-sampled. Microstepping and interlacing allow
sampling on scales of 1/2 or 1/3 of a pixel (see Section
\ref{sec_implement}). The layout of the four detectors and their
numbering scheme is shown in Figure \ref{wfcam_focal_plane}. The
detectors cannot be butted because of their packaging. Instead they
are spaced at $94\%$ of the detector width, as shown. Four exposures
are required to survey a contiguous area, referred to as a `Tile' (see
Section \ref{sec_general_implement}). As indicated in Figure
\ref{wfcam_focal_plane}, each detector is rotated by $90^{\circ}$ with
respect to its neighbours. The pixel $x, y$ coordinate convention is
indicated for the detectors.

Each detector is divided into quadrants, and each quadrant is divided
into eight channels of $128\times1024$ pixels. The channel packing
relative to the $xy$ axes is illustrated in Figure
\ref{wfcam_focal_plane}. The packing rotates through $90^{\circ}$ in
moving to an adjacent quadrant. This channel packing arrangement is
visible for detector 4 in Figure \ref{flaws}.

The optical path is illustrated in \citet{casali06}. The design
involves a new $f/9$ secondary, which produces an intermediate focus
5.7m above the primary. A tertiary mirror returns the beam upward to
the downward--looking focal plane. For the purposes of this paper, an
important element of the design is a field lens, which gives rise to
two types of artifact in the data (Section \ref{sec_data_flaws}).  The
field lens images the primary mirror to a cold stop inside the
cryostat, which baffles stray thermal emission. The lens lies near the
intermediate focus, and forms the end of the barrel of the
instrument. There is an auxiliary lens mounted at the centre and just
above the field lens. The purpose of the auxiliary lens is to raise
the autoguider focal plane, and to change the $f$-ratio to achieve the
required pixel scale. The autoguider is located on the optical axis,
as shown.  It was necessary to raise the guider CCD above the focal
plane because of the packaging of the Hawaii-II detectors.

WFCAM has eight filter housings. Because the focal plane vignettes the
beam, to minimise throughput loss the filters are stored vertically,
rather than in a wheel. Consequently the detectors are exposed to
unfiltered light when the filter is changed. Currently, the standard
filter complement includes the broadband set $ZYJHK$, and two
narrow-band filters, $H_2$ 1-0 S1 \& Br$\gamma$. The eighth filter
housing is blanked for darks. The WFCAM $J$, $H$ and $K$ bandpasses
follow the specification of the Mauna Kea filter set
\citep{tokunaga02}.  The WFCAM $Z$ filter has an effective wavelength
of $0.87\mu$m and the WFCAM $Y$ filter fills the gap between $Z$ and
$J$. The system throughput, above-atmosphere-to-detector, is close to
$20\%$ at $J$, slightly higher at $H$ and $K$, and slightly lower at
$Z$ and $Y$. The $ZYJHK$ photometric system of the instrument has been
characterised by \citet{hewett06}, who provide the total system
response curves for each broad band, as well as colour equations to
translate to 2MASS $JHK$ (see also Section
\ref{sec_photo_calib}) and SDSS $z$, and synthetic colours of a
wide variety of stars, galaxies and quasars.

UKIDSS observations are made in one of two read modes. In correlated
double sampling (CDS), after a detector reset, an initial read is
made, followed by an exposure of the desired length, followed by a
final read. The resulting output is the difference of the two
reads. In non-destructive read (NDR) mode, after the detector reset,
several reads sample the build up of charge throughout the exposure,
and the counts are established by a linear fit. CDS has higher
read-noise $\sim30e^-$, but the likelihood of electronic pick-up due
to the reduced read activity is smaller than with NDR. CDS is used for
all UKIDSS $JHK$ observations, which use shorter integrations because
of the high background. NDR has lower read-noise $\sim20e^-$, but is
affected more by amplifier glow (although at present, this is well
removed by dark subtraction). Also, although measurements currently
indicate that detector responses are linear \citep{irwin06}, if the
need for linearisation should arise in the future, then this
correction is a more challenging computational problem than with CDS.
NDR is used for all $ZY$ observations, which use longer exposures, to
achieve sky-noise limited performance.

The detectors are of PACE manufacture. The frames exhibit four
characteristics that are unwelcome: pixel covariance, curtaining,
persistence, and cross-talk. {\em Pixel covariance} is a result of
inter-pixel capacitance. The result is that the pixel to pixel
variance is reduced by a factor of about 1.2, and that the readout
gain (photons per ADU) is overestimated by this factor when computed
in the traditional way using photon--noise transfer curves. In this
paper, where appropriate, e.g. in computing the system throughput, we
have compensated for the effect. {\em Curtaining} is a rapidly varying
constant bias offset along each `line', where the pattern repeats in
each quadrant (rotating in the same manner as the channel
packing). The term `line' here means 1024 pixels along the direction
of the short axis of the channels.  The pattern fluctuates with a root
mean square (RMS) of 10e$^-$ to 15e$^-$ and is completely random; no
two detector frames are ever affected in the same way.  {\em
Persistence} is the effect of image latency whereby counts from a
bright source are not completely removed after a detector reset.
Artificial images of sources can also occur due to {\em cross-talk}
between the detector channels within a quadrant. Persistence and
cross-talk are described in Section
\ref{sec_pipeline} and all effects are discussed at greater 
length by \citet{irwin06}.

\section{The UKIDSS 2-year Plan}
\label{sec_2year_plan}

The design and layout of the surveys are set out in
\citet{lawrence06} which is the reference work for details of the
final (7-year) goals of the UKIDSS programme. These goals are
summarised in Table \ref{tab_survey_final} in terms of areas,
filters, number of passes, and (final summed) depth. All depths quoted
in this paper are in the Vega photometric system and correspond to
the total brightness of a point source for which the summed flux in a
$2''$ diameter aperture has, on average, S/N=5. The original
designed depths were based on a first estimate of sensitivities in
each band, made in 2001 and later revised brighter. It is apparent from
analysis of the EDR dataset that the achieved sensitivities are indeed
shallower than the original estimates, by an average of 0.2  mag. In 
light of this, all target depths have been adjusted by 0.2 mag and
these are the values quoted here and in \citet{lawrence06}. The
overestimate of the detector quantum efficiency, due to pixel
covariance (Section \ref{sec_wfcam}), is the main contribution to the
overestimate of the instrument throughput.

In each of the shallow surveys, the depth in a select filter ($J$ for
LAS, $K$ for GCS, GPS) is built up in two or three passes, in order to
provide proper motions and variability information. Thus any field is
observed in the full filter complement at one epoch and observed in
the select filter only at one or two other epochs.  The repeat passes are
separated by a minimum of two years, to provide a sufficient baseline
for measurement of proper motions. For scheduling reasons, in some
areas the first epoch is with the full filter complement, and in other
areas the first epoch is with the select filter.

\begin{table}
\centering
\begin{tabular}{@{}crlcc@{}}
\hline
Survey & Area (deg$^2$) & Filter & No. passes & Depth \\ 
 \hline
         &      & $Y$ & 1 & 20.3 \\
  LAS    & 4028 & $J$ & 2 & 19.9 \\
         &      & $H$ & 1 & 18.6 \\
         &      & $K$ & 1 & 18.2 \\
\hline
         & 1868 & $J$ & 1 & 19.9 \\
  GPS    & 1868 & $H$ & 1 & 19.0 \\
         & 1868 & $K$ & 3 & 19.0 \\
         &  300 & $H_2$ & 3 & ... \\
\hline
         &      & $Z$ & 1 & 20.4 \\
         &      & $Y$ & 1 & 20.3 \\
  GCS    &  1067 & $J$ & 1 & 19.5 \\
         &      & $H$ & 1 & 18.6 \\
         &      & $K$ & 2 & 18.6 \\
\hline
         &   35 & $J$ & multiple & 22.3 \\
  DXS    &    5 & $H$ & multiple & 21.8 \\
         &   35 & $K$ & multiple & 20.8 \\
\hline
         &      & $J$ & multiple & 24.8 \\
  UDS    & 0.77 & $H$ & multiple & 23.8 \\
         &      & $K$ & multiple & 22.8 \\
 \hline
\end{tabular}
\caption{Summary of the final (7-year) goals of the UKIDSS
programme. Depth is in the Vega system and corresponds to
the total brightness of a point source
for which the flux in a $2''$ diameter aperture has S/N=5. The GPS
depths refer to uncrowded fields.}
\label{tab_survey_final}
\end{table}

The surveys are currently focused on completing a set of goals for the
first two years of operations, hereafter referred to as the 2-year
plan. In this section we describe the 2-year plan, which provides the
context of the EDR and explains, for example, why some areas are
covered in only a single band. In the 2-year plan the LAS and GCS
are accelerated relative to the other surveys.

\begin{figure*}
\epsfxsize=167mm
{\hfill
\epsfbox{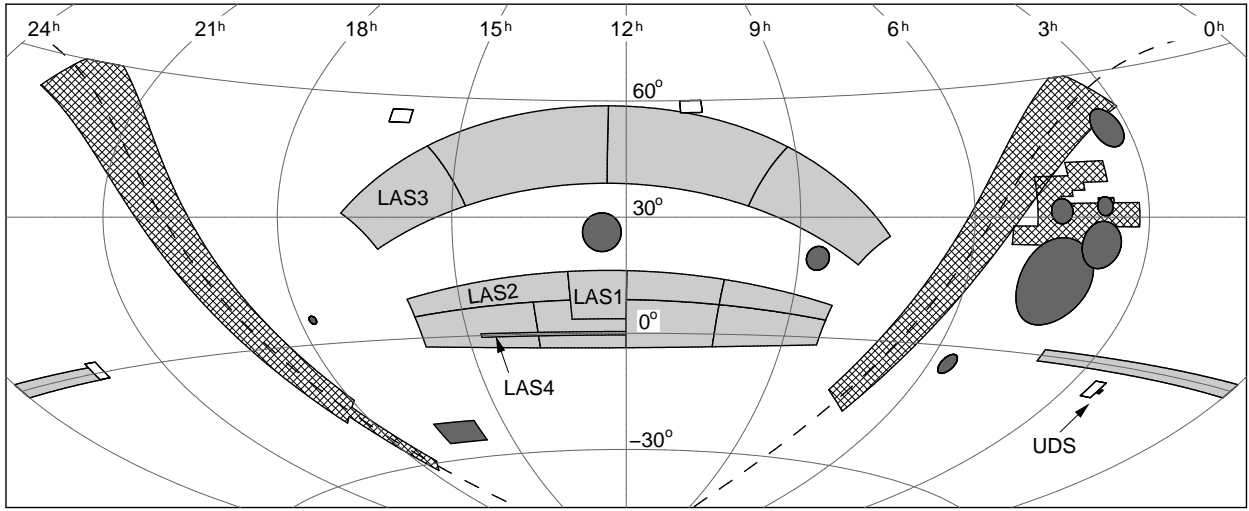}
\hfill}
\caption{The UKIDSS 2 year plan survey areas showing the LAS (solid
light grey), GPS (cross-hatched), GCS (solid dark grey), DXS (empty
squares) and UDS (as labelled, lying alongside the western-most DXS
field). The dashed line indicates the galactic plane. The LAS is shown
here divided into its constituent projects (see Section
\ref{sec_general_implement}), the EDR containing data in LAS1-4.
Note that in the 2-year plan the LAS northern block receives $J$
only coverage and most of the GPS areas shown receive $K$ only (see
text).}
\label{ukidss_survey_areas}
\end{figure*}

Table \ref{tab_survey_defns} summarises the relevant details of the
2-year plan for each survey, listing area to be surveyed, filter,
target depth and default integration time, for the shallow surveys,
for good conditions. As explained in Section \ref{sec_implement}, in
cases of mediocre seeing, bright sky or thin cirrus, observations in
some fields are still undertaken, but the integration times are
increased to ensure the target depth is achieved. The geometry of the
target areas in the 2-year plan is shown in Figure
\ref{ukidss_survey_areas}. In what follows, we provide additional
details of the 2-year plan, describing each survey in turn. Section
\ref{sec_edr} describes the data content of the EDR.

\begin{table}
\centering
\begin{tabular}{@{}lccccr@{}}
\hline
Survey & Area (deg$^2$) & Filter & Depth & t$_{\rm int}$ \\
                                    \hline
       & 2120 & $Y$   & 20.3 & 40s  \\
 LAS   & 3120 & $J$   & 19.5 & 40s  \\
       & 2120 & $H$   & 18.6 & 40s  \\
       & 2120 & $K$   & 18.2 & 40s  \\           
                                    \hline
       &  338 & $J$   & 19.9 & 80s  \\          
 GPS   &  338 & $H$   & 19.0 & 80s  \\ 
       & 1868 & $K$   & 18.2 & 40s  \\    
       &  300 & $H_2$ & ...  & 150s \\
                                    \hline
       & 237  & $Z$   & 20.4 & 40s  \\
       & 237  & $Y$   & 20.3 & 40s  \\
 GCS   & 237  & $J$   & 19.5 & 40s  \\
       & 237  & $H$   & 18.6 & 40s  \\
       & 606  & $K$   & 18.2 & 40s  \\
                                    \hline
       & 12.6 & $J$   & 22.3 &      \\
 DXS   & 12.6 & $K$   & 20.8 &      \\
                                    \hline
       & 0.77 & $J$   & 23.8 &      \\
 UDS   & 0.77 & $K$   & 22.8 &      \\
                                                       \hline
\end{tabular}
\caption{Summary of the goals of the UKIDSS 2-year plan.  Depth is in
  the Vega system and corresponds to the total brightness of a point
  source for which the flux in a $2''$ diameter aperture has S/N=5.
  The GPS depths refer to uncrowded fields. For the shallow surveys
  the default integration time for normal 
  conditions is listed. Some fields are observed in mediocre conditions, with
  longer integration times, as detailed in Section
  \ref{sec_implement}.}
\label{tab_survey_defns}
\end{table}

\subsection{LAS 2-year plan}

The LAS is made up of 3 blocks, visible in
Figure \ref{ukidss_survey_areas}. All three blocks lie within the SDSS
footprint. In the southern galactic hemisphere there is an equatorial
stripe, Block 1, of area 212 deg$^2$. In the northern galactic
hemisphere, there is an equatorial block, Block 2, and a block at
higher declinations, Block 3, both of area 1908 deg$^2$. The LAS
2-year plan is to survey Block 1 and Block 2 in a single pass in
$YJHK$. In addition, depending on conditions (Section
\ref{sec_implement}), part of Block 3, an estimated 1000 deg$^2$, will
be covered in a single pass in $J$ only. For scheduling purposes each
block has been subdivided into regions. The EDR contains data from the
regions marked LAS1, LAS2, LAS3 \& LAS4 (see Section \ref{sec_edr}).

\subsection{GPS 2-year plan}

The GPS will cover the band of galactic latitude $|{\rm
b}|<5^{\circ}$, that lies within the declination range
$-20^{\circ}<\delta<+60^{\circ}$. This translates to two blocks
defined by the galactic longitude ranges $15^{\circ}<{\rm
l}<107^{\circ}$ and $142^{\circ}<{\rm l}<230^{\circ}$. In addition,
there will be a thin extension into the galactic Centre, defined by
$|{\rm b}|<2^{\circ}$, $-2^{\circ}<{\rm l}<15^{\circ}$, as well as
imaging of the $\sim 300$ deg$^2$ Taurus-Auriga-Perseus star-formation
complex. The geometry of Taurus-Auriga-Perseus is shown by the
irregular cross-hatched area in Figure \ref{ukidss_survey_areas} which
follows the CO gas map of \citet{ungerechts87}

The fields for the first two years will concentrate on the region of
the inner Galaxy, close to the plane.  The region $|{\rm
b}|<1^{\circ}$ in the longitude range $-2^{\circ}<{\rm l}<107^{\circ}$
will be imaged in $JHK$. This includes the area covered by the
northern half of the GLIMPSE survey \citep{churchwell02} using the
Spitzer Space Telescope. Also, the block covering $30^{\circ}<{\rm
l}<45^{\circ}$ will be imaged in $JHK$, over the remainder of the full
thickness, $|{\rm b}|<5^{\circ}$. The whole of the remainder of the
GPS area will be surveyed in $K$ only, excluding Taurus-Auriga-Perseus
which will be imaged in $H_2+JHK$ in clear weather, or $H_2+K$ in thin
cirrus conditions. This ensures a 2 year baseline between $H_2+K$
observations (at least) of the complex over the full 7 year plan.  For
more information and details of updates, we refer the reader to the
GPS web site located at {\em
http://star-www.herts.ac.uk/$\sim$pwl/UKIDSS\_GPS.html}.

\subsection{GCS 2-year plan}

The GCS will observe the 10 star clusters listed in
Table \ref{tab_gcs_fields}. These are ranked by priority, which gives
an indication of the sequence in which they will be completed. In the
2-year plan, five of the clusters will be completed in all filters, as
well as the central regions of three other clusters, as listed in the
table. In addition, the two remaining clusters will be covered in $K$
only.

For the GCS, the EDR contains data  from the
Sco and Coma-Ber clusters. For more information and details
of updates, we refer the reader to the GCS web site located at {\em
http://www.roe.ac.uk/$\sim$nch/gcs/}.

\begin{table}
\centering
\begin{tabular}{@{}ccccc@{}}
\hline
Field & Type & RA, Dec & Filters & Area (deg$^2$) \\
\hline
IC 4665 & OC &	17:46,+05:43 & $ZYJHK$ & 3.1\\
Pleiades & OC &	03:47,+24:07 & $ZYJHK$ & 79  \\
Alpha Per & OC & 03:22,+48:37 &	$ZYJHK$ & 50  \\
Praesepe & OC &	08:40,+19:40 & $ZYJHK$ & 28  \\
Tau.-Aur. &	SFA & 04:30,+25:00 & $ZYJHK$ & cen. 24  \\
Orion &	SFA & 05:29,$-$02:36 &	$ZYJHK$ & cen. 16  \\
Sco &	SFA & 16:10,$-$23:00 &	$ZYJHK$ & cen. 75 \\
Per-OB2 & SFA &	03:45,+32:17 & $ZYJHK$ & 12.6 \\
Hyades & OC &	04:27,+15:52 & $K$ only & 291  \\
Coma-Ber & OC &	12:25,+26:06 & $K$ only & 78.5  \\
\hline
\end{tabular}
\caption{The GCS fields, with details of coverage in the 2-year
  plan. Types are open cluster (OC) or star formation association (SFA)}
\label{tab_gcs_fields}
\end{table}

\subsection{DXS 2-year plan}

The DXS comprises 12 contiguous Tiles in each of four fields; XMM-LSS
(2$^{\rm h}25^{\rm m}$,$-4^{\circ}30'$), ELAIS N1 (16$^{\rm h}10^{\rm
m}$,+54$^{\circ}00'$), the Lockman Hole (10$^{\rm h}57^{\rm
m}$,+57$^{\circ}40'$) and VIMOS 4 (22$^{\rm h}17^{\rm
m}$,+0$^{\circ}20'$). The 2-year plan is to reach full depth in four
Tiles in $J$ and $K$ in each of the four fields. 
Rather than build up depth over all four Tiles, the strategy is
to achieve target depth in one Tile before moving on to the next.

\subsection{UDS 2-year plan}

The UDS is a single $0.8$ deg$^2$ Tile, centred on (2$^{\rm
h}18^{\rm m}$,$-5^{\circ}10'$), chosen to coincide with the
Subaru/XMM-Newton Deep Survey field. This field lies just westward of
the XMM-LSS field in the DXS.  The 2-year plan is to
cover the Tile to $K=22.8$, i.e. full depth, and to $J=23.8$,
i.e. one magnitude short of full depth. 

\section{Survey observational implementation}
\label{sec_implement}

In this section we set out relevant details of the observational
implementation that are needed for understanding the characteristics
of the data. The observing procedures were adjusted in the interval
between 05A and 05B, and relevant differences are detailed in the
appropriate sections.

\subsection{General details}
\label{sec_general_implement}

Each survey is divided into `projects'.  Projects are a means of
segregating observations for administrative purposes.  In the shallow
surveys, projects segregate observations by area, for example, the LAS
is divided into several projects, the EDR containing data in the
projects LAS1-4 (see Figure \ref{ukidss_survey_areas}).
In the deep surveys, projects simply provide a convenient way of
dividing up observations between semesters. 

Observations are packaged in `minimum schedulable blocks' (MSBs).
MSBs typically last one hour and are designed to be self contained in
the sense that the sky frames for sky subtraction may be formed from
the data itself. Each MSB has its own set of criteria for the
conditions under which it may be executed. At the summit the UKIRT
observing management program automatically schedules MSBs
according to their observability (i.e. hour angle and airmass) and
the current conditions (transparency, seeing, $J$-band sky brightness).

The basic unit of the surveys is the stack `{\em multiframe}' (see
Section \ref{sec_wsa_arch}), which is the group of four images, one
for each detector, formed by averaging the set of exposures made at
the same base position within a given MSB. The region being surveyed
is tessellated by these stack multiframes. The smallest contiguous
region that can be surveyed is a Tile, requiring four base positions
to fill in a solid angle of 0.77 deg$^2$.

Descriptions of the implementation of each survey employ the
following observing terminology, defined by UKIRT, which we use with
appropriate capitalisation to distinguish from the general term:
\begin{itemize}

\item {\em Exposure} - An Exposure is a single complete
cycle of the detector readout mode, over a time specified as the
Exposure Time, $t_{\rm exp}$ (referred to by the {\tt expTime} attribute
in the WSA).

\item {\em Integration} - An Integration is a sequence of one or more
Exposures, that are averaged by the data acquisition system, resulting
in a single image being written to disk. An Integration is therefore
the basic observational product. For all UKIDSS observations,
Integrations comprise single Exposures.

\item {\em Offset} - An Offset is an accurate shift in the telescope
pointing between two Integrations. Integer-pixel Offsets are used for
all UKIDSS Integrations, except in the case of Microstepping (see
below). At UKIRT, Offsets smaller than $10''$ are fast, so all Offsets
used in UKIDSS MSBs are smaller than this. Also, frames Offset by
$<10''$ may be combined without regard to WFCAM's variable pixel
scale (see Section \ref{sec_astrom_calib}). Larger Offsets required
rebinning of the data.

\item {\em Microstep} - A Microstep is a special case of an Offset,
where the step in X and Y has the size of either $n+1/2$ or $n \pm 1/3$
pixels, $n$ being an integer. These two cases are referred to as
$2\times2$ and $3\times3$ Microstepping hereafter.

\end{itemize}

A variety of Offset patterns are used in the surveys, resulting in three
different types of frame in the archive as follows:
\begin{itemize}

\item {\em Normal frames} - A Normal frame corresponds to an Integration. 

\item {\em Interleave frames} - A set of Microstepped frames (four or
  nine, for $2\times2$ or $3\times3$ Microstepping, respectively) are
  interleaved onto the sub-pixel grid, to form an Interleave frame
  (abbreviated `Leav' elsewhere in this paper and in the WSA).

\item {\em Stack or Interleave-Stack frames} - Where Microstepping has
  not been used, frames taken at different Offsets are registered, and
  averaged to form a `Stack' frame. Where Microstepping has been used,
  the Interleave frames are registered and averaged to form an
  Interleave-Stack frame (abbreviated `Leavstack' elsewhere in this
  paper and in the WSA).

\end{itemize}

Note that because the zero point for a frame relates counts per
second to magnitudes, and because averaging is used at all points in
combining frames, only $t_{\rm exp}$ is needed to calibrate any frame
(rather than the total Integration time, $t_{\rm tot}$).

\subsection{Shallow surveys: LAS, GCS \& GPS}

The choice of $t_{\rm exp}$, whether to Microstep, and the number of
Offsets, involves consideration of the relative importance of
sampling, overheads, bad pixels and read-noise relative to sky
noise. These reasons motivate the different implementation schemes
detailed below. Between them, the three shallow surveys use three
different implementation schemes, numbered 1 to 3 in Table
\ref{tab_obs_design}. For each scheme, the table lists successively the
scheme number, the filters to which the details apply, the Exposure
time, the Microstepping (if used), the number of Offset positions
(other than Microsteps) and the total Integration time
$t_{\rm tot}$ making up the stacked frame.

\begin{table}
\centering
\begin{tabular}{@{}lccccc@{}}
\hline
Scheme  & Filter & $t_{\rm exp}$ & $\mu$-step? & Offsets & $t_{\rm tot}$ \\
(Survey) &       & (s)          &         &         &   (s)        \\
 \hline
1 (LAS, GCS) & $ZY$  & 20s & no           & 2-pt & 40 \\
  & $JHK$ & 5s  & $2 \times 2$ & 2-pt & 40 \\ \hline
2 (LAS) & $Y$  & 20s & no           & 4-pt & 80 \\
  & $JHK$ & 10s & $2 \times 2$ & 2-pt & 80 \\ \hline
3 (GPS) & $JH$  &  5s & $2 \times 2$ & 4-pt & 80 \\
 & $K$   &  5s & $2 \times 2$ & 2-pt & 40 \\ \hline
4 (DXS) & $JK$  & 5s  & $2\times2$   & 25-pt & 500 \\ \hline
5 (DXS) & $JK$  & 10s & $2\times2$   & 16-pt & 640 \\ \hline
6 (UDS) & $JK$  & 10s & $3\times3$   & 9-pt  & 810 \\ \hline
\end{tabular}
\caption{The different implementation schemes used for the shallow
  surveys, LAS, GCS \& GPS}
\label{tab_obs_design}
\end{table}
   
\subsubsection{LAS implementation}
\label{sec_las_implement}

LAS observations are carried out in the filters $Y$, $J$, $H$, and
$K$. The goal is to reach uniform depth in all fields. The MSBs
comprise observations of a number of Tiles and are of two types,
where the filters used are either the pair $YJ$ or the pair $HK$.  If
an area has been covered by one pair, every effort is made to cover
the same area with the other pair, within a month. Nevertheless, this
is not a strict requirement. In order to make the most efficient use
of the observing time, a wide range of conditions is used. The choice
of MSB to execute depends on whether the conditions fall into one of
the three categories listed in Table \ref{tab_las_obscon}, defined
by photometricity, seeing, and $J$-band sky brightness. Since these
are the conditions pertaining at the start of the MSB, the actual
conditions may be better or worse. The final criteria which
define any data release are set at the QC stage (Section
\ref{sec_qc}).

\begin{table}
\centering
\begin{tabular}{@{}ccccc@{}}
\hline
Category  & Photometric & Seeing & $J$ sky \\
          &             &        & mag/arcsec$^2$ \\ 
\hline
  A          &     Y       & $<1.0\arcsec$ & $>15.7$ \\
  B          &     Y       & $<1.4\arcsec$ & $>15.0$ \\
  C          &  thin       & $<1.0\arcsec$ & $>15.7$ \\ \hline
  \end{tabular}
\caption{The different categories of observing conditions 
allowed for LAS observations}
\label{tab_las_obscon}
\end{table}
   
In 05A, observations were made in the four regions marked LAS1, LAS2,
LAS3 and LAS4 in Figure \ref{ukidss_survey_areas}. The combinations of
filters, implementation strategy (Table \ref{tab_obs_design}), and
conditions category (Table \ref{tab_las_obscon}), used for each region
are as follows: LAS1 and LAS4 ($YJHK$, 1, A), LAS2 ($YJHK$, 2, B),
LAS3 ($J$, 2, C), which translates as follows: If conditions are
photometric, with good seeing and dark sky, regions LAS1 and LAS4 are
observed in all filters. If conditions are photometric, but either
the seeing is poor or the sky is bright, or both, Integration times
are doubled and LAS2 is observed in all filters. If there is thin
cirrus but the seeing is good and the sky dark, Integration times are
doubled and the first epoch $J$-band observations in LAS3 are
made. Because calibration is made using 2MASS stars in the frames
themselves (Section \ref{sec_photo_calib}), the extinction by cloud in
these observations is corrected for. However, this correction accounts
only for the average extinction over the frame, not spatial variation in
extinction, hence the accuracy of the photometry will not be as good
as under photometric conditions.  This remains to be quantified.

\subsubsection{GPS implementation}
\label{sec_gps_implement}

GPS observations are carried out in the filters $J$, $H$ and $K$,
using strategy 3 (Table \ref{tab_obs_design}). Depending on
conditions, two different types of MSB are executed.

\begin{itemize}

\item[1)] In photometric conditions, good seeing $<0.8\arcsec$ at $K$
  and $J$ sky brightness $>15.5$ mag/arcsec$^2$, a single Tile is
  observed in filter order $J$, $H$ then $K$, providing near-simultaneous
  observations of the same area.

\item[2)] In conditions of thin cirrus and good seeing $<0.8\arcsec$
  at $K$, first-epoch $K$-band observations of four Tiles are executed.

\end{itemize}

\subsubsection{GCS implementation}
\label{sec_gcs_implement}

GCS observations are carried out in the filters $Z$, $Y$, $J$, $H$ and $K$
using scheme 1 (Table \ref{tab_obs_design}), in seeing
conditions $<1.0\arcsec$. Depending on
the transparency, two different types of MSB are executed.

\begin{itemize}

\item[1)] In photometric conditions the full filter set is
  observed. For $J$ sky brightness $>15.7$ mag/arcsec$^2$, MSBs
  containing observations in $ZYJ$ may be executed. There is no sky
  brightness constraint on MSBs containing observations in $HK$.
  
\item[2)] In conditions of thin cirrus first-epoch $K$-band
observations are executed.

\end{itemize}

\subsection{Deep surveys: DXS \& UDS}

The two deep surveys employ similar strategies. An MSB creates the
necessary data to produce four stack multiframes covering a Tile in a
single filter. Depth is built by repeating the MSB starting from a
slightly different base position. For example, for the UDS, the
centres are produced with a Gaussian distribution about the global
field centre with a FWHM of $20''$, and truncated at a maximum offset
of $2'$.

\subsubsection{DXS implementation}
\label{sec_dxs_implement}

DXS observations are carried out in the filters $J$ and $K$. An MSB
comprises observations in one of the filters, of a complete
Tile. Implementation scheme 4 (Table \ref{tab_obs_design}) was used in
05A and implementation scheme 5 was used in 05B. The reason for the
change to longer Integration times was to reduce overheads. For the
EDR, DXS observations required seeing $<1.1''$ and transparency
conditions of thin cirrus or better.

\subsubsection{UDS implementation}
\label{sec_uds_implement}

UDS observations are carried out in the filters $J$ and $K$. An MSB
comprises observations in one of the filters, of the complete UDS
Tile. Implementation scheme 6 (Table \ref{tab_obs_design}) was
used. For the EDR the required observing conditions for $J$ were
seeing $<0.85''$, sky brightness $>16.0$ mag/arcsec$^2$, and for $K$,
seeing $<0.75\arcsec$ with no sky brightness limit. Required
transparency was thin cirrus or better, for all observations.

\section{Pipeline and calibration}
\label{sec_pipeline}

Pipeline processing and archiving for all WFCAM data, including the
UKIDSS surveys, is provided by the VISTA Data Flow System Project
(VDFS), which is a collaboration between Queen Mary University of
London, Cambridge, and Edinburgh. The UKIDSS consortium interacts
closely with the VDFS team and has primary responsibility for the
final QC stage in producing the UKIDSS products.

At the summit there is a data stream for each detector. Thus, every
Integration results in a single FITS file being written to each of
four LTO tapes. After transfer to CASU\footnote{The web page {\em
http://www.ast.cam.ac.uk/$\sim$wfcam} provides access to the WFCAM raw data
and includes summaries of the progress of the observing and data
reduction for the surveys.} but prior to pipeline ingest, the four
FITS files of an Integration are combined into a single
multi-extension FITS (MEF) file. A MEF file has one header per
extension and a primary header that contains general observation
information together with the data acquisition protocols used, which
trigger the appropriate pipeline processing components. These are
described in this section. 

The output of the pipeline is a set of reduced uncalibrated Normal
frames and photometrically and astrometrically calibrated
Stack/Leavstack frames, as well as catalogues of images detected in
the Stack/Leavstack frames. Merging of catalogues across bands is
undertaken in the archive. We refer the reader to \citet{irwin06} for
further information on all details contained in this section.

\subsection{Pixel data processing}

The pipeline processes data on a night-by-night basis. The data files
fall into one of three categories; science frames (including
calibration standards), (dawn) twilight flatfields, and 
darks. The first pipeline step creates a set of master darks by
stacking frames segregated by Exposure time and read mode. The
appropriate master dark is then subtracted from all
frames. Aside from hot pixels, the dark current is very low for the
WFCAM detectors and the main effect of dark subtraction is to remove
the majority of the reset anomaly\footnote{This is an additive
component varying in amplitude in a ramp--like manner across a frame,
as a consequence of the decaying capacitative behaviour of the
detector, and the time delay between the reset and the first read, as
a function of position.}.

The data are then flattened by the appropriate master twilight flat,
for the particular filter. Master twilight flats are updated
at least monthly. These are used in preference to dark--sky
flats which can be corrupted by thermal emission and fringing
(although the latter is at a very low level in the WFCAM
detectors). We note in passing that detectors 1 and 4 are respectively
the most and least sensitive, with a quantum efficiency ratio of $\sim
1.3$ in $Z$, falling to $\sim 1.2$ in $K$. This variation is accounted
for by normalising each detector's flatfield by the median of the
counts in all four detectors.

The next stage is to subtract the bias curtaining, described in
Section \ref{sec_wfcam}. The curtaining pattern is mirrored, rotated,
in each quadrant. Therefore, the varying bias level may be established
by appropriate filtering of the data in $4 \times 1024$
pixels. Following this, a running median--filtered sky--correction
frame is created and subtracted for removal of residuals, including
images due to dirt and marks on the field lens, either from scattered
light or thermal emission. The term `sky correction' means subtraction
of residuals relative to flat sky, i.e. the average sky count is
preserved.

Different versions of the pipeline were used to process 05A (LAS, GPS,
GCS, some DXS) and 05B (UDS, some DXS) data. The 05B version has
improved sky correction and a module to correct cross-talk
images. Presently, the pipeline makes no correction for
persistence. This is currently being characterised and a later version
may include removal of the effect.

If appropriate, the frames are then interleaved (Section
\ref{sec_general_implement}). The final stage is the creation of a
Stack or Leavstack frame by weighted averaging.  Each frame is
weighted by the product of the inverse of its noise variance and the
confidence map. The confidence map is derived principally from the
master flatfield and is normalised to have a median of unity, but
additionally dead pixels, pixels with poor sensitivity and pixels with
unpredictable levels (e.g. some hot pixels) are given zero confidence.
For the DXS and UDS, where multiple visits are made, the Leavstack
frames are themselves stacked, but at a later stage (see Section
\ref{sec_edr}).

\subsection{Catalogues}
\label{sec_catalogues}

A source extraction routine based on the methodology of
\citet{irwin85} is run on the stacked images. Sources are identified
with groups of four or more connected pixels that all lie more than
$1.5\sigma$ above the local background sky level ($1.25\sigma$ for
frames interleaved to $0.2\arcsec$/pixel or $0.133\arcsec$/pixel). A
global background following algorithm \citep{irwin84} is used to track
the varying sky level over each frame, on a scale of $25.6''$ (i.e.
64, 128 or 192 pixels for no, $2\times2$ or $3\times3$ Microstepping).
This scale represents a compromise between being small enough to
follow rapidly varying background and large enough to sample
sufficient sky pixels outside objects. The initial background value in
each $25.6\arcsec\times25.6\arcsec$ block is computed using an
iterative k-sigma clipped median estimator, after removing bad
pixels. The resulting array of block values is then filtered to remove
outliers due, for example, to bright stars. The filtered coarse grid
is then bi-linearly interpolated back to the original pixel grid.

A set of parameters is measured for each detected object. These are
listed in Table \ref{tab_detection_table}. Fluxes are corrected for
the flatfield error, caused by the variable pixel scale across the
field of view (Section \ref{sec_astrom_calib}). Errors are computed
from photon count, read-noise and local background error. A number of
parameters that will be included in later versions of the pipeline,
including PSF magnitudes and Sersic-profile fits are also listed. In
addition, a variety of quality control parameters are recorded for
each stacked detector frame, including average stellar ellipticity and
seeing, as well as the sky brightness and the photometric zero
point. In deriving object parameters, the current version of the
pipeline does not make use of the weight information in the confidence
map.  The measured parameters include a series of 13 circular-aperture
fluxes of fixed aperture size, listed as {\tt
aperFlux[1-13]}. Apertures are `soft-edged' in the sense that flux in
pixels bisected by the aperture boundary is scaled according to the
fraction of the pixel enclosed. All aperture fluxes (and the
corresponding magnitudes {\tt aperMag[1-13]}) in this set have the
aperture correction applied, as determined from bright stars in the
field. This means that for point sources, these aperture magnitudes
approximate total magnitudes so that for typical seeing, the parameter
{\tt aperMag3} provides the most accurate estimate of the total
magnitude. There are three further aperture fluxes, the Petrosian,
Kron and Hall fluxes, where the aperture size is defined from
measurement of the surface--brightness profile. These three fluxes are
not corrected for flux outside the aperture. Merged objects are
deblended in a manner which avoids double counting of flux.

\subsection{Astrometric calibration}
\label{sec_astrom_calib}

Astrometric calibration incorporates a cubic radial distortion term,
i.e. where the relation between $r_{\rm true}$, the true on-sky
angular distance from the optical axis, and $r$, the distance measured
in a WFCAM image, takes the form
\begin{equation}
\label{eq_rad_distn}
r_{\rm true}=r + k r^3 \, .
\end{equation}
The constant $k$ has a small wavelength dependence. Measured values at
$H$ and $K$ are $-50$/radian$^2$ and $-53$/radian$^2$, respectively,
for $r$ in radians. Higher order distortion terms are insignificant,
compared to the specified absolute systematic accuracy of
$<0.1\arcsec$ RMS. Calibration is achieved by a fit to the coordinates
of objects in the 2MASS point source catalogue, solving for distortion
parameterised by $k$, and a six-parameter linear transformation;
\begin{equation}
\label{eq_linear_trans}
x'={\rm a}x+{\rm b}y+{\rm c} \quad ; \quad y'={\rm d}x+{\rm e}y+{\rm d} \, ,
\end{equation}
which allows for translation, scaling, rotation and shear. Since 2MASS
astrometry is based on Tycho 2 and is in the International Celestial
Reference System (ICRS), so too is WFCAM astrometry. This does not
require an Equinox to be specified and for most practical purposes this
is (almost) equivalent to Equinox 2000 in FK5\footnote{MEF files have the
FITS keyword RADECSYS = `ICRS' set. For backwards compatibility with 
existing software, EQUINOX = 2000.0 is left as is.}.
Once computed, the astrometric solution is written into the catalogue
MEF files and back-propagated to the stacked and component image MEF files.

\subsection{Photometric calibration}
\label{sec_photo_calib}

The procedures for photometric calibration of UKIDSS data are still
being refined. The specified absolute systematic accuracy is $0.02$ mag
in each band. Initially it was expected that calibration would be
achieved through setting up fields of WFCAM standards, which
themselves would be calibrated to UKIRT Faint Standards. Instead the
EDR data have been calibrated using 2MASS stars in the frames
themselves, applying colour equations to convert 2MASS photometry to
the WFCAM system \citep[see][for details of the WFCAM
filters]{hewett06}. Further improvements of this method are
anticipated, and it now appears likely that the specified accuracy can
be achieved through this route.  In this section we describe the
photometric calibration of the EDR, which summarises the current
status of our programme to use 2MASS stars to calibrate UKIDSS fields.

The motivation for basing the calibration on 2MASS includes:
\begin{itemize}
\item The 2MASS global calibration is accurate to better than 2\%
across the entire sky \citep{nikolaev00}.
\item In each WFCAM field there is always an abundance of unsaturated stars,
with 2MASS errors $<0.1$ mag (typically 60-1000, depending
on galactic latitude).
\item Measuring the calibrators at exactly the same time as the targets
enables photometric calibration even during non-photometric conditions
(accuracy limited by the uniformity of the cloud cover).
\end{itemize}

To calibrate the frames, the first step is to identify unsaturated
WFCAM sources that appear in the 2MASS Point Source Catalogue, with
2MASS errors $<0.1$ mag. The 2MASS photometry is transformed into the
WFCAM $ZYJHK$ system using the following empirically--derived colour
equations:
\begin{eqnarray}
Z_{\rm WFCAM} &=& J_{\rm 2MASS}+ 0.95(J_{\rm 2MASS}-H_{\rm 2MASS}) \nn \\
Y_{\rm WFCAM} &=& J_{\rm 2MASS}+ 0.50(J_{\rm 2MASS}-H_{\rm 2MASS}) \nn \\
J_{\rm WFCAM} &=& J_{\rm 2MASS}-0.075(J_{\rm 2MASS}-H_{\rm 2MASS}) \nn \\
H_{\rm WFCAM} &=& H_{\rm 2MASS}+0.04(J_{\rm 2MASS}-H_{\rm 2MASS})-0.04 \nn\\
K_{\rm WFCAM} &=& K_{\rm 2MASS}-0.015(J_{\rm 2MASS}-K_{\rm 2MASS}) \nn
\end{eqnarray}
A zero point is then computed for each star, as
\begin{equation}
\label{eq_zp}
ZP_i = m_i + 2.5 \, {\rm log}_{10}(N_i/t_{\rm exp})
\end{equation}
where $N_i$ is the parameter {\tt aperFlux3} for the star, i.e. the
total counts in ADU of source $i$, estimated by aperture correction of
the summed counts in an aperture of radius $1''$ and
additionally corrected for extinction, using a default extinction of
0.05 mag per unit airmass (assumed for all nights and all filters). The
photometric zeropoint for a given stacked detector frame is then
simply the median zeropoint over all unsaturated stars. The extinction
correction would not be needed simply for calibrating the data, but
allows the zero points to be inter-compared, to provide a measure of
photometricity. Note that since counts in interleaved frame pixels are
the counts in the full WFCAM pixels from the progenitor Normal frames,
when comparing zero points between different frames, a correction of
$2.5\log_{10}(4)$ or $2.5\log_{10}(9)$ must be subtracted from the zero
points of frames Microstepped by $2\times2$ or $3\times3$
respectively.

The accuracy of the astrometry and photometry in the EDR data is
quantified in Section \ref{sec_qc}.  

\section{Data artifacts}
\label{sec_data_flaws}

\begin{figure*}
\epsfxsize=170mm
{\hfill
\epsfbox{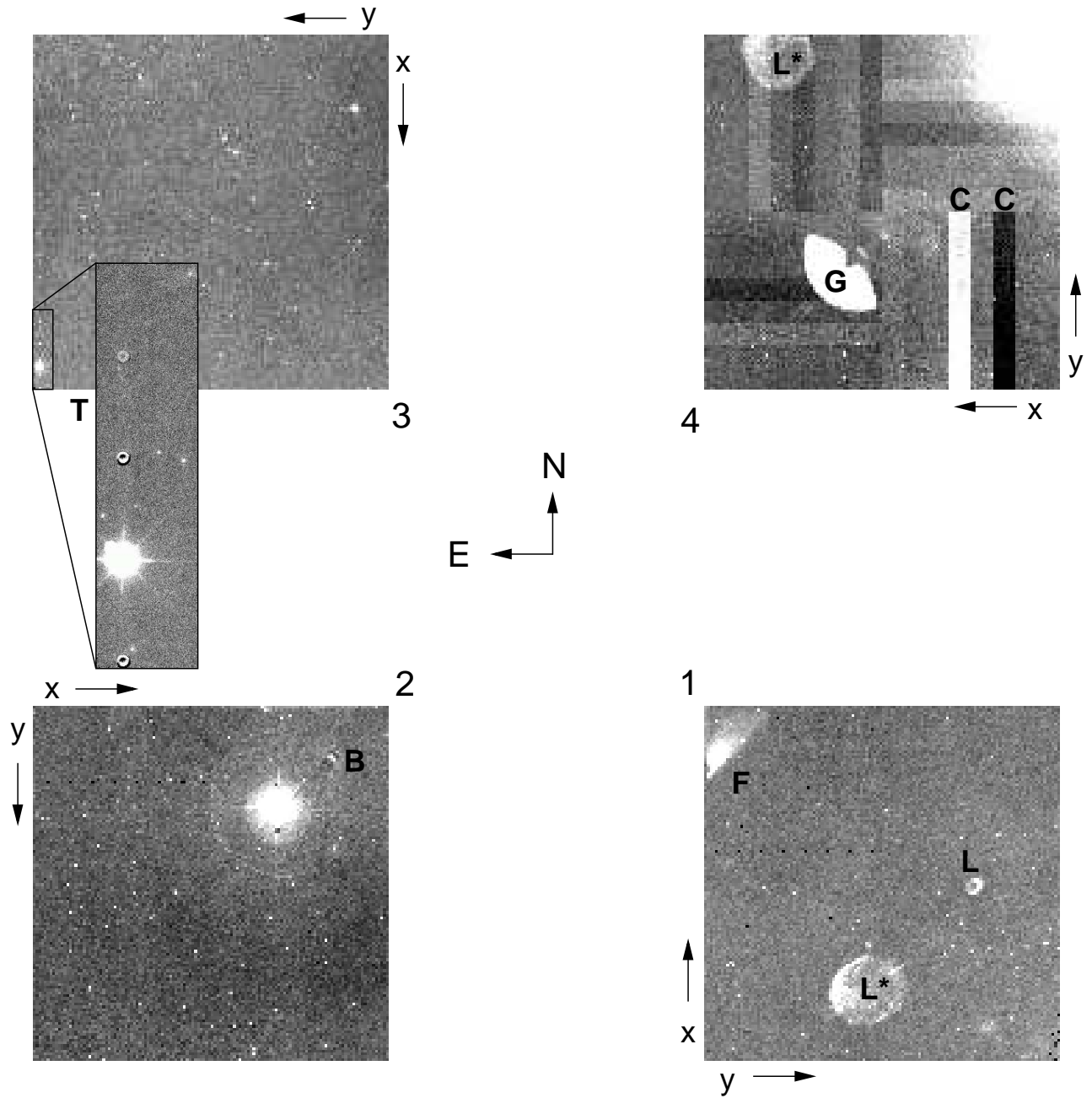}
\hfill}
\caption{WFCAM focal plane showing composite of various data artifacts
that can arise in WFCAM data, described in detail in the text. As
marked, North is up and East to the left. Data artifacts key: F -
flare persistence image that appears after filter changes, T -
cross-talk image sequence associated with bright star, C - the two
worst channels with variable bias level, G - bright ghost from
moonlight shining on the field lens, L - images caused by marks or
dust on the field lens (L* are marks left by the lens
handling jig), B - `bow-tie' feature in the halos of bright stars.}
\label{flaws}
\end{figure*}

There are a variety of artifacts in the images which compromise the
quality of the data in a number of ways. In particular, they give rise
to false detections, with the consequence that a small fraction of the
entries in the catalogues are for objects that have apparently
interesting colours, but are not real. For reference purposes we
catalogue the various artifacts here, as a prelude to the next
section, where the QC procedures are described. Figure \ref{flaws}
provides a montage illustrating the various effects.

\begin{flushleft}
{\em Persistence images} 
\end{flushleft}

Persistence images of a saturated star can appear over several
subsequent frames and are not currently treated in the pipeline. The
brightness of the persistence image is a function of source count
rate, wavelength (i.e. filter), and number of reset/reads.
Persistence images are characteristically the size of the saturated
part of their progenitor image. Images persisting across a number of
frames will follow any Offset pattern employed and in such cases the
effect usually stacks out.

The flare image marked `F' in Figure \ref{flaws} is also a result of
persistence and is often visible in 05A data in the first frame taken
after a filter change (because the detector has been exposed to
unfiltered light and consequently saturated; see Section
\ref{sec_wfcam}). For the 05B data, a short sequence of dark frames
following each filter change was implemented, to flush persistence
images. In these data, the flare image is thus largely removed.

\begin{flushleft}
{\em Cross-talk images} 
\end{flushleft}

The zoom image in Figure \ref{flaws} of a portion of detector 3
illustrates the problem of cross-talk. Here, a sequence of spurious
images can be seen, which are caused by the bright star. In general,
cross-talk images appear in some or all of the 7 other channels of the
quadrant in which the star is located. The images are always offset
across the channel width by an integer multiple of 128 pixels (256 or
384 pixels for $2\times 2$ or $3 \times 3$ Microstepping,
respectively). The appearance of the cross-talk images depends on the
counts in the star, and, for bright stars, the shape matches the
spatial derivative of the profile of the star. Therefore, for bright
unsaturated stars the cross-talk images have a half-moon appearance,
while for saturated stars, where the counts at the centre of the image
are constant at the saturation value, the cross-talk images have the
curious washer appearance shown. Cross-talk images in the channels
adjacent to the star have features at the level of $\sim1\%$ of the
flux difference between adjacent pixels in the star, dropping to lower
levels further from the source.

Cross-talk images are particularly pernicious because they appear in
the same location relative to the bright star, for all filters.
Therefore the detections are matched across bands and the objects
appear in the catalogues as entries with peculiar colours.

\begin{flushleft}
{\em Channel bias offsets} 
\end{flushleft}

In a few percent of frames the counts in any channel may be offset
high or low by a few counts, or more rarely by tens of
counts. Because of the lower sky counts, the effect is more often
visible in the shorter wavelength bands $Z$, $Y$ and $J$. Two
channels in detector 4, marked `C' in the figure, are particularly
prone to this variable bias level. The offset can vary from one frame
to the next and therefore is not effectively removed by subtraction
of a dark frame. Very occasionally all the channels in a detector are
afflicted, such that the entire channel pattern is visible, as in the
case of detector 4 in the illustration.

Channel bias offsets are particularly troublesome in interleaved data,
since every fourth pixel (or ninth in $3 \times 3$ Microstepping) is
offset in level, resulting in a `bed-of-nails' pattern. In the worst
cases, the bias offsets confuse the object detection algorithm,
resulting in a large number of spurious sources in the channel. We
have attempted to identify the worst cases and removed the frame and
catalogues from the release database. Both hardware and software
solutions to the problem of channel bias offsets are being pursued.

\begin{flushleft}
{\em Bright moon ghosts} 
\end{flushleft}

In some circumstances, apparently when the moon is shining directly
onto the field lens, bright ghost images occur, such as the image
marked `G' in detector 4 in Figure \ref{flaws}. The appearance of the
ghosts depends on the angle of inclination of the moon relative to the
optical axis. The most common case, of the type illustrated, occurs
when the moon is less than $30^{\circ}$ from the target. At larger
angles, highly elongated arcs sometimes appear. The exact cause of
these images is under investigation. It is possible that the ghosts
arise from reflection off the mounting arrangement of the autoguider
auxiliary lens at the centre of the field lens.

We have attempted to identify all the brightest examples of moon
ghosts, and remove the affected frames and accompanying image
catalogues. Moon ghosts occur both in 05A and 05B data, affecting
nearly $3\%$ of the fields observed. For future WFCAM campaigns,
fields within $30^{\circ}$ of the moon will be avoided. Additionally a
baffle extending above the field lens will be installed, which will
further reduce the problem.

\begin{flushleft}
{\em Images of marks or dirt on the field lens} 
\end{flushleft}

Because the field lens (Section \ref{sec_wfcam}) is located close to
an intermediate focus, marks or dirt on the lens, are visible in some
frames. The three worst examples are marked `L'/`L*' in Figure
\ref{flaws}. The two large blotches, marked `L*', correspond to the
locations of support pads used in the installation of the lens. These
are particularly visible in scattered light when the moon shines
directly on the field lens (the case shown is a particularly bright
example). Their visibility has been much reduced since the underside
of the lens was cleaned between semesters 05A and 05B. There are also
numerous smaller elliptical-ring images, that are caused primarily by
dust on the upper surface of the lens. These images are visible both
in scattered light and in thermal radiation. They are more prevalent
in the earliest data, until more frequent cleaning of the upper
surface was introduced. As a rule, when illumination from the moon is
fairly constant, these spurious images are successfully removed by the
pipeline, at the stage of sky correction. However, the problem is more
difficult to treat if the intensity is variable, for example as the
moon transits into or out of the dome slit. Avoiding observations
close to the moon, and installation of the field-lens baffle, should
both act to reduce the problem of images of marks or dirt on the field
lens.

\begin{flushleft}
{\em Diffraction pattern for bright stars} 
\end{flushleft}

The image of a bright star in detector 2 is shown in Figure
\ref{flaws}. The diffraction pattern shows 8 spikes, from two spiders,
one supporting the secondary, and the other the guider auxiliary lens,
above the field lens. The spikes can extend to large angles, $\sim
20^{\prime}$, and therefore may be visible in some frames where the
bright star itself is not visible. There is a low-level feature in the
point spread function, marked `B' in Figure \ref{flaws}, which has the
shape of two bow-ties. Similarly this feature can be visible in some
frames, where the star itself is outside the field of view.

\section{UKIDSS Data Quality Control}
\label{sec_qc}

This section is divided into two. The first part describes the UKIDSS
QC procedures and summarises the fraction of data rejected. The
second part quantifies the accuracy of the astrometric and photometric
calibration. The areal coverage and depth of each survey are provided
in the next section.

\subsection{QC procedures and data quality}

The output of the pipeline is a set of Normal, Leav and
Stack/Leavstack frames and catalogues of detections in the
Stack/Leavstack frames. These data are ingested by the archive, where
a process of curation results in seamless catalogues of objects
matched across bands. In collaboration with WFAU we are developing a
set of QC procedures, specifically for UKIDSS data, that are applied
at various points in the ingest and curation processes. The aim of the
QC is to eliminate data that are either corrupt (i.e. where something
went wrong in the data-train and the data are meaningless), or bad
(i.e. where the data are unusable, because of a bright moon ghost, for
example), and then to apply a set of cuts that define `survey
quality' using a number of measured parameters, such as seeing and
depth. The definitions for each survey of `survey quality' are still
being refined, with the aim of finalising a set of rules that will be
applied rigidly to DR1. We have used the EDR to develop the rules. For
the EDR we have defined the cuts (see below) by inspecting the
distributions of the relevant parameters and clipping the
tails. Bearing in mind the poorer image quality of the 05A data, we
anticipate that the rules applied to DR1 will be more stringent than
those applied to the EDR.

First, a set of book-keeping checks are made to identify corrupt
data resulting from glitches in the data-train. For example, if the
wrong information is written to the headers, a set of files may have
been processed with the wrong algorithm. For a particular field,
observed with a particular filter, the checks include verifying that
the correct complement of files and catalogues exist, that data are
present for each detector, and that the header parameters are
meaningful. Depending on the nature of the problem, the entire
complement of files and catalogues is eliminated either for a
detector, or for the full set of four detectors.

Three checks are made to identify bad data. Every stack frame is
inspected visually for cases of bright moon ghosts, trailing
(i.e. where guiding was lost), poor sky correction, or particularly bad
cases of channel bias offsets. A second check for cases of poor sky
correction is made using the attribute {\tt skySubScale}. This
is the scaling of the correction frame made in the
sky-correction step. A check is made of all normal frames, for cases
where the parameter is unusually high or low. A third check for bad
data is made by identifying frames with large numbers of false
objects, where the detection algorithm has been confused. This can
arise in cases of large channel bias offsets and also sometimes where
the sky correction is poor. The check is made by searching the merged
catalogues for frames containing unusually large numbers of objects
detected in one band only.

The cataloguing procedure results in a set of parameters that are
useful for characterising the quality of the data, for each detector of
a stack multiframe. Listing the parameters by the name stored in the
WSA table {\tt Multiframedetector} (Section \ref{sec_wsa}), these are:
{\tt seeing} (the average image FWHM of detected stellar sources),
{\tt avStellarEll} (the average ellipticity of detected stellar
sources), {\tt skyLevel} (the average counts per pixel in blank sky),
{\tt skyNoise} ($\sigma_{sky}$, a robust measure of the standard
deviation of counts in the sky), and {\tt PhotZPCat} ($m_0$, the
photometric zero point). From these, we compute the $5\sigma$ detection
limit for a point source $m_d$, as 5 times the standard deviation of the
counts in an aperture, corrected for flux outside the aperture, by the
aperture correction $m_{\rm ap}$
\be
\label{eq_depth}
m_d=m_0-2.5{\rm log}_{10}(5\sigma_{\rm sky}(1.2N)^{1/2}/t_{\rm
  exp})-m_{\rm ap} \ee 
where $N$ is the number of pixels in the
aperture, and the factor 1.2 accounts for the covariance between
pixels.

\begin{figure*}
\epsfxsize=175mm
{\hfill
\epsfbox{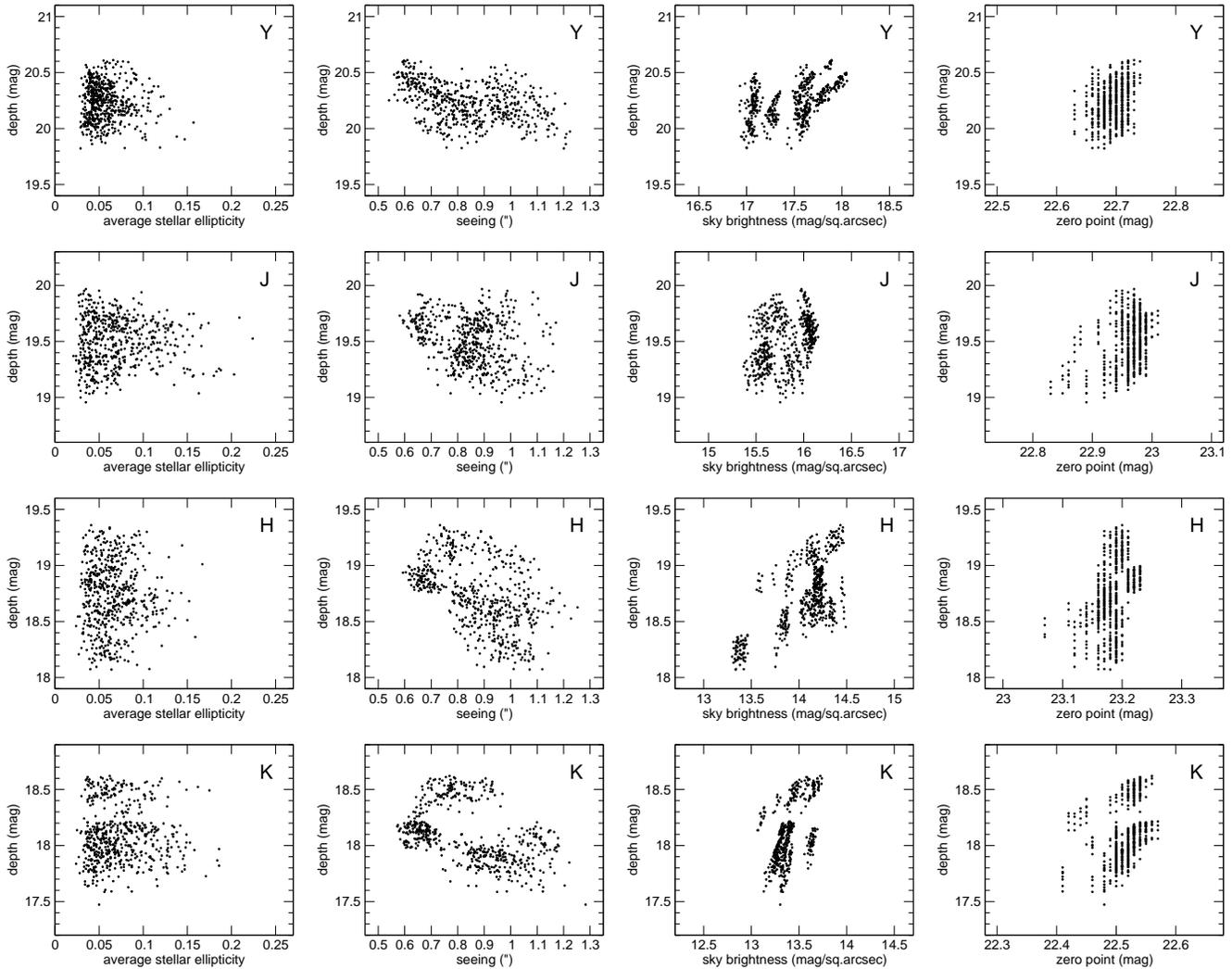}
\hfill}
\caption{LAS data quality, plotting depth from equation
(\ref{eq_depth}) versus average stellar ellipticity, seeing, sky
brightness and zero point in $Y$, $J$, $H$ \& $K$. Each point
represents one detector frame in the EDR database. Note that a few
points lie beyond the seeing limit of $1.2''$ since the cut was applied
to the average multiframe seeing, not individual detector frames.}
\label{las_qc_plots}
\end{figure*}

The interplay between these various parameters is illustrated in Figure
\ref{las_qc_plots}, which plots depth $m_d$ against ellipticity,
seeing, sky brightness and zero point, for the LAS data in the EDR
database, i.e. all LAS fields with data in all four filters which pass
the QC cuts (defined below). A number of points are worth noting. The
plots show the expected anti-correlations between depth and seeing, and
depth and sky brightness (most clearly seen in the $H$-band plot). The
effect of thin cloud is also noticeable, e.g. in the $K$ band, in the
correlation between depth and zero point. In most cases, the average
ellipticity is $<0.1$ with a tail to larger values. Large
ellipticities may arise in fields where the guide star is very faint
and guiding was lost. This occurred occasionally in the LAS, where, of
all the surveys, the surface density of suitable bright guide stars is
lowest. We have since adjusted our tiling strategy to minimise this
problem.

For the EDR, we applied seeing limits for each survey, marginally
greater than the limits specified for the observations, as follows;
DXS ($1.25\arcsec$), LAS ($1.2\arcsec$), GCS ($1.2\arcsec$), GPS
($1.0\arcsec$) \& UDS ($J$ $0.9\arcsec$, $K$ $0.8\arcsec$). These were
applied to the average measured value for the four detectors of each
stack multiframe, i.e. the whole multiframe was accepted or rejected on
this measure. A cut on ellipticity $<0.25$ was applied to all
data. For most fields a cut was made on zero point at a level of 0.2
mag extinction from the modal (i.e. photometric) value. In other words,
a number of fields have been calibrated through thin cloud. This
figure of 0.2 mag is under review, pending an analysis of the
uniformity of photometric calibration under such conditions. For the
DXS and UDS fields and the first-epoch single-band observations in
the shallow surveys, extinction up to 0.3 mag was permitted, although
conditions this bad are rare in the data. Finally, to ensure
uniformity, a cut on depth was made at 0.4 mag brighter than the modal
value for each filter in each survey.

%

Discounting corrupt data removed by the book-keeping stage of QC,
the quantity of sub-standard data rejected by the QC procedure over 
the whole of the EDR (i.e. the EDR+ database) as a fraction of 
each survey is as follows: 
LAS 23.0\%, GPS 27.3\%, GCS 25.9\%, DXS 19.6\% \& UDS 0.0\%.  The
corrupted data approximately account for an extra third of these
fractions per survey. Of the corrupted data, $> 95\%$ will be
recovered by re-processing in a later release.

The proportion of data rejected, nearly 25\% overall, is rather large,
but this figure should fall to below 10\% in the future. Of the 25\%,
about 10\% represents repeat observations, because the first
observations were unsatisfactory. Revised procedures for identifying
deteriorating conditions and interrupting the observations will reduce
this figure. A further substantial fraction, about 7\%, can be
attributed to scattered light or bright ghosts (and associated sky
gradients), when observing near the moon. These issues will be dealt
with in more detail in the paper accompanying the First Data Release,
but have now (July 2006) been solved. A further 3\% of frames are
rejected for a variety of reasons, and the final 5\% are frames which
are outside the quality control cuts, for seeing, cloud etc., where
seeing accounts for the majority. The implementation strategy allows
for observations under a very wide range of conditions (seeing, sky
brightness, cirrus) in order to make the most efficient use of the
observing time available. Consequently the figure of 5\% of frames
observed in poor conditions is viewed as acceptable.

\subsection{Calibration accuracy}

The results of internal and external checks of the accuracy of the
astrometric and photometric calibration are summarised here.

\subsubsection{Astrometric calibration}

\begin{flushleft}
{\em Internal astrometric accuracy}
\end{flushleft}

\begin{figure}
\epsfxsize=70mm
{\hfill
\epsfbox{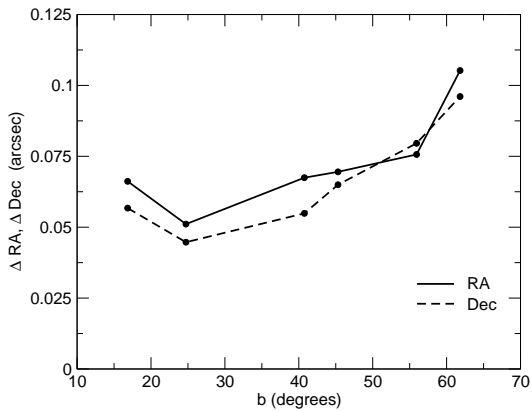}
\hfill}
\caption{Astrometric consistency of duplicate objects found in
overlapping detector frame edges in the LAS and GCS. Plotted is the
RMS/$\sqrt{2}$ of the distribution of the median difference in RA or
Dec of duplicate stellar-like objects (with $12 \le J \le 18$) per
detector overlap area.}
\label{las_int_ast}
\end{figure}

The astrometric calibration, using 2MASS (Section \ref{sec_pipeline}),
will be accurate to the extent that the functional form of the
transformation applied is an accurate representation of the mapping
between WFCAM pixel coordinates and equatorial coordinates, and there
are sufficient 2MASS stars in the field. An internal check of the
astrometric accuracy is possible by considering the consistency of
the measured positions of objects observed twice, because they lie in
overlapping edges of detector frames in adjacent fields. This will be
a stringent test, since any systematic errors are likely to be largest
at the frame edges.  Figure \ref{las_int_ast} shows the result of this
analysis for the $J$ band, for the LAS and GCS. For each overlap
region, the median value of the difference in RA and in Dec was
computed for all stellar-like objects detected twice. The RMS
(divided by $\sqrt{2}$ to give the systematic error on a single
measurement) of this median difference, for all the overlap regions,
is plotted against galactic latitude.  At b$\sim 20^{\circ}$, the
error in RA and Dec is $\sim50$mas in each coordinate, rising to
$\sim100$mas at b$\sim 60^{\circ}$, where the fields contain fewer
2MASS stars for computing the solution.

\begin{flushleft}
{\em External astrometric accuracy}
\end{flushleft}

\begin{figure*}
\epsfysize=120mm
{\hfill
\epsfbox{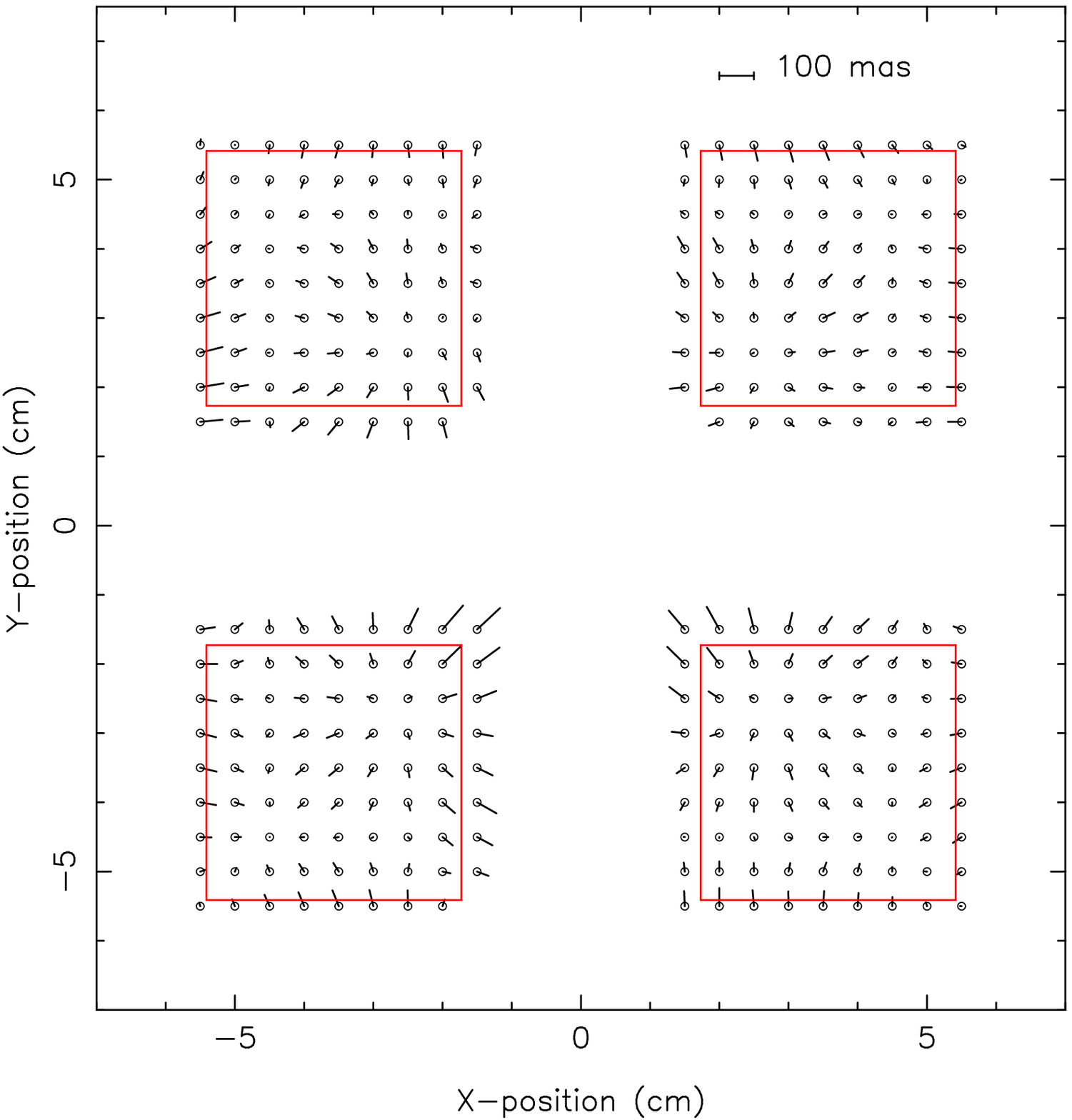}
\hfill}
\caption{Distribution of astrometric systematic errors across WFCAM's
 field of view, calculated by averaging the residuals from 2MASS of
 WFCAM observations of $\sim$150,000 stars per detector.}
\label{2mass_astrom}
\end{figure*}

If we assume that any systematic errors in the 2MASS astrometry itself
will average out over many fields, we can measure the systematic
errors in the astrometry around the WFCAM focal plane, by computing
the differences between calibrated WFCAM positions and 2MASS
positions, and averaging over observations of many fields at the same
position on the detector. The result for $\sim$1100 fields,
representing a week of EDR $JHK$ observations, is shown in Figure
\ref{2mass_astrom}. The plot shows the average residuals for all stars
detected at S/N$>10$. The size of the averaged RA and Dec residuals is
23mas RMS on each coordinate. With the current calibration scheme,
this represents the limiting accuracy achievable, for the case of a
field containing many 2MASS stars. The scatter for individual stars
contributing to the plot is 80mas in each coordinate, which represents
the random errors on the 2MASS coordinates. This gives an indication
of the accuracy achievable in a field containing few 2MASS
stars. Subject to further analysis, to confirm that the systematics
plotted in the figure are stable, they will be corrected for in future
releases.

\subsubsection{Photometric calibration}

\begin{flushleft}
{\em Internal photometric accuracy}
\end{flushleft}

\begin{figure}
\epsfxsize=85mm
{\hfill
\epsfbox{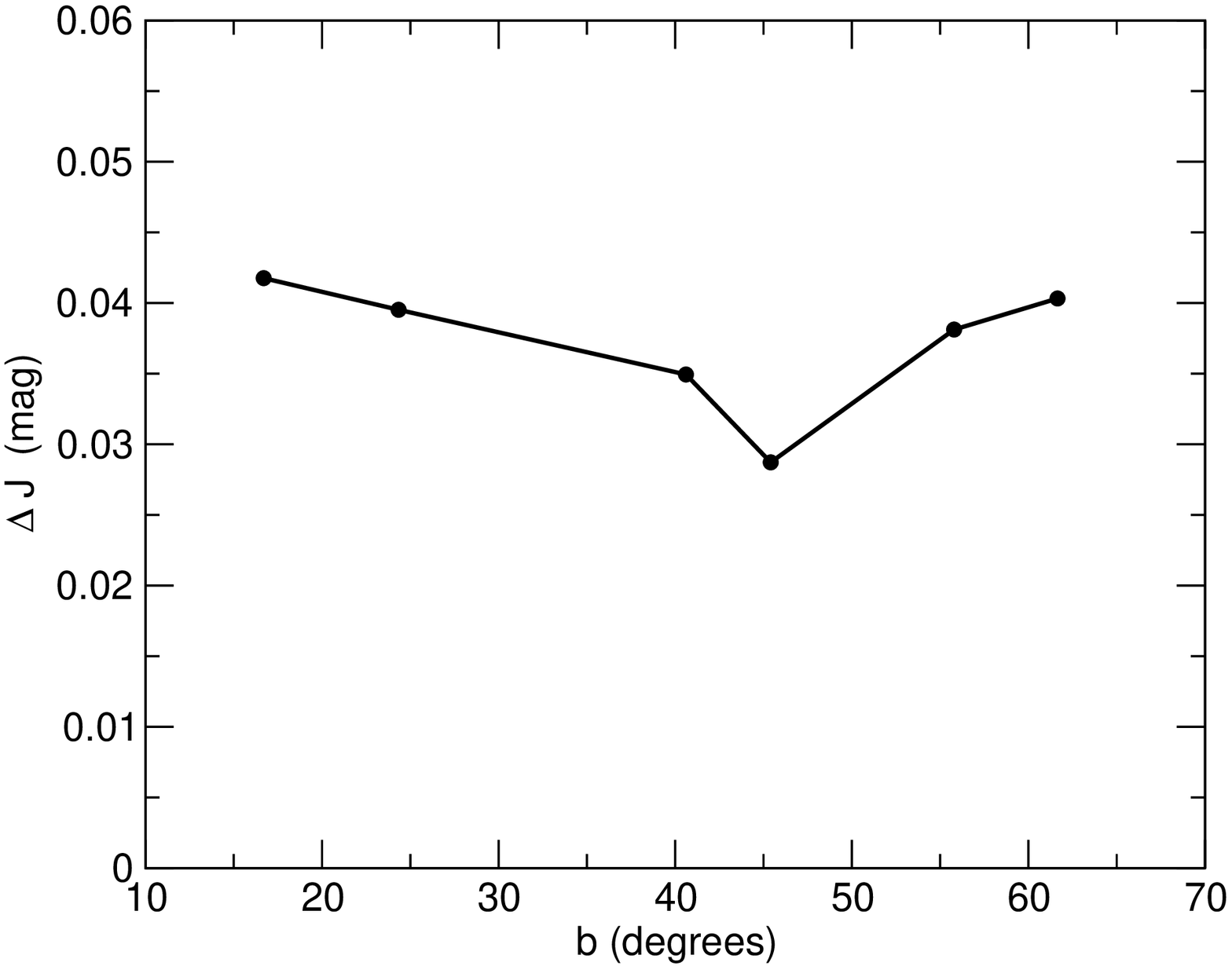}
\hfill}
\caption{Photometric consistency of duplicate objects found in
overlapping detector frame edges in the LAS and GCS. Plotted is the
RMS/$\sqrt{2}$ of the distribution of the median difference in $J$
band magnitude of duplicate stellar-like objects (with $12 \le J \le
18$) per detector overlap area.}
\label{J_phot_vs_b}
\end{figure}

A similar check using the overlap regions has been made to estimate
the internal photometric accuracy. The results are shown for the $J$
band in Figure \ref{J_phot_vs_b}. As before, the median offset in
photometry of stellar-like objects is computed for each overlap
region and the RMS/$\sqrt{2}$ computed for all overlap
regions. The results are plotted as a function of galactic
latitude. In this case there is no visible
trend with galactic latitude, and the average error is slightly less
than 0.04 mag. Because the measurements are made at the edges of each
detector, this probably overestimates the internal error representative
of the whole field.

\begin{flushleft}
{\em External photometric accuracy}
\end{flushleft}

Because of the large number of 2MASS stars in each field (Section
\ref{sec_catalogues}), the accuracy of the photometry will be set by
systematic errors in the 2MASS photometry and the accuracy of the
colour transformations. For the 05A and 05B semesters, hourly
observations were made of fields containing UKIRT faint standards
\citep{hawarden01}, in all the broadband filters. These observations
have been calibrated using 2MASS in exactly the same way as our UKIDSS
fields. Therefore comparison of the 2MASS-calibrated photometry of the
UKIRT faint standards, against the published values provides an
external check of the accuracy.

\begin{figure}
\epsfxsize=85mm
{\hfill
\epsfbox{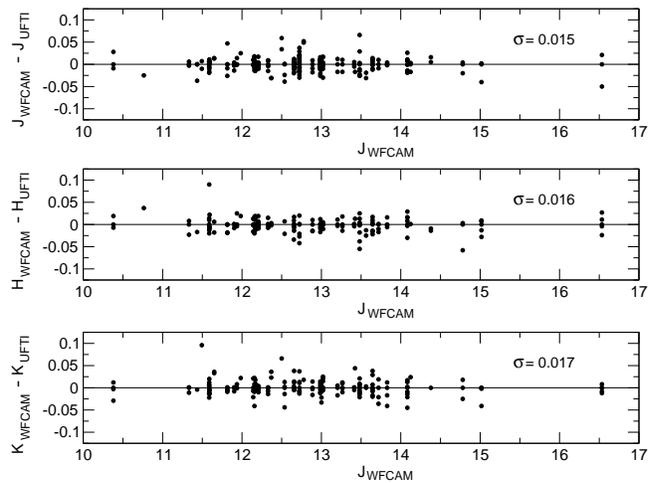}
\hfill}
\caption{Difference between 2MASS calibrated WFCAM magnitudes and
UFTI magnitudes of 46 UKIRT faint standard stars in $J$, $H$ \& $K$ as a
function of WFCAM $J$ magnitude.}
\label{jhk_photo_calib}
\end{figure}

The results for 600 measurements of 46 standard stars, measured on 19
photometric nights, for $J$, $H$ and $K$ are plotted in Figure \ref
{jhk_photo_calib}.  We find that the RMS scatter for these
measurements is less than 2\%, which is the UKIDSS requirement for
photometric accuracy \citep[see][]{hodgkin06}.

The quality of the $Z$ and $Y$ calibration has not yet been quantified.

\section{The contents of the EDR}
\label{sec_edr}

This section provides an overview of the contents of the EDR,
including maps of the areas covered and details of the depths
reached. The three shallow surveys LAS, GPS \& GCS are described 
first, followed by the two deep surveys, DXS \& UDS.

\subsection{Shallow survey data}
\label{sec_shallow_contents}

LAS, GPS and GCS data in the EDR were observed exclusively in semester
05A, between the dates 15th May 2005 and 20th June 2005. Table
\ref{tab_shallow_coverage} provides details of the area covered by
each survey. The first column lists the survey, and the second column
provides the area covered in the EDR database, i.e. fields with
observations in all filters for that survey. The next five columns
list the area covered in the EDR+ database, by filter. The final
column provides the summed area of fields covered by any filter.  Note
the large coverage in $J$ in the LAS and in $K$ in the GPS, because
of the inclusion of first-epoch observations in the particular filter.

\begin{table}
\centering
\begin{tabular}{@{}cccccccc@{}}
\hline
Survey & EDR  & \multicolumn{6}{c}{EDR+ DB} \\
       &  DB  &  $Z$ &  $Y$ &  $J$ &  $H$ &  $K$ & any \\
\hline
LAS    & 27.3 &   -  & 33.5 & 89.4 & 57.3 & 51.0 & 114.2 \\
GPS    & 14.9 &   -  &   -  & 16.8 & 17.8 & 57.9 &  59.7 \\
GCS    &  7.0 & 13.4 & 14.4 & 16.0 & 32.1 & 29.7 &  38.3 \\ \hline
sum    & 49.2 & 13.4 & 47.9 &122.2 &107.2 &138.6 & 212.2 \\
\hline 
\end{tabular}
\caption{Coverage of the shallow surveys (deg$^2$) in the EDR and EDR+
  databases.}
\label{tab_shallow_coverage}
\end{table}

The depths reached in the three shallow surveys are summarised, by
filter, in Table \ref{tab_shallow_depth}. The quantity listed is the
$5\sigma$ depth for a point source, computed using equation
\ref{eq_depth}. Two columns are provided for each survey. The first
column lists the median value of the depth measured for all detector
fields in the EDR database. The second column lists the same quantity,
but restricted to fields in which the seeing was
$<1''$. Because the image quality improved between semesters 05A
and 05B, we expect that most of the data in future releases will have
seeing $<1''$, so quantities in the second column are
expected to be more representative of `survey quality'. Nevertheless
the differences are quite small.

The values quoted for the GPS deserve comment. The total integration
times for the $J$ and $H$ bands are 80s, twice as long as for the LAS
and GCS in these bands, while the total integration time in the $K$
band, 40s, is the same as for the LAS and GCS (see Table
\ref{tab_obs_design}). Taking the integration times into account, the
GPS data are less sensitive than the LAS and GCS data by 0.4 mag on
average, over the $JHK$ bands. These values are for the EDR database
and therefore are for fields within $1^{\circ}$ of the plane. The
lower sensitivity is attributed to unresolved stars in the background
increasing the sky noise. Because nearly all the data have seeing
$<1''$, the depths quoted in the two columns are identical for the
GPS.

\begin{table}
\centering
\begin{tabular}{@{}ccccccc@{}}
\hline
       & \multicolumn{2}{c}{LAS} & \multicolumn{2}{c}{GPS}  &
\multicolumn{2}{c}{GCS} \\
Filter & all & $<1\arcsec$ & all & $<1\arcsec$ & all & $<1\arcsec$ \\
 \hline
$Z$ &  $-$  &  $-$  &  $-$  &  $-$  & 20.29  &  20.40 \\
$Y$ & 20.23 & 20.25 &  $-$  &  $-$  & 19.92  &  20.05 \\
$J$ & 19.52 & 19.51 & 19.51 & 19.51 & 19.40  &  19.46 \\
$H$ & 18.73 & 18.79 & 18.67 & 18.67 & 18.79  &  18.80 \\
$K$ & 18.06 & 18.12 & 17.74 & 17.74 & 18.12  &  18.12 \\
\hline
\end{tabular}
\caption{The median $5\sigma$ point source depth by filter,
in the EDR database, for the three shallow surveys, LAS, GPS \& GCS. 
The two columns for each survey are for all seeing and
just those data with seeing $<1''$.}
\label{tab_shallow_depth}
\end{table}

\begin{figure*}
\epsfxsize=160mm
{\hfill
\epsfbox{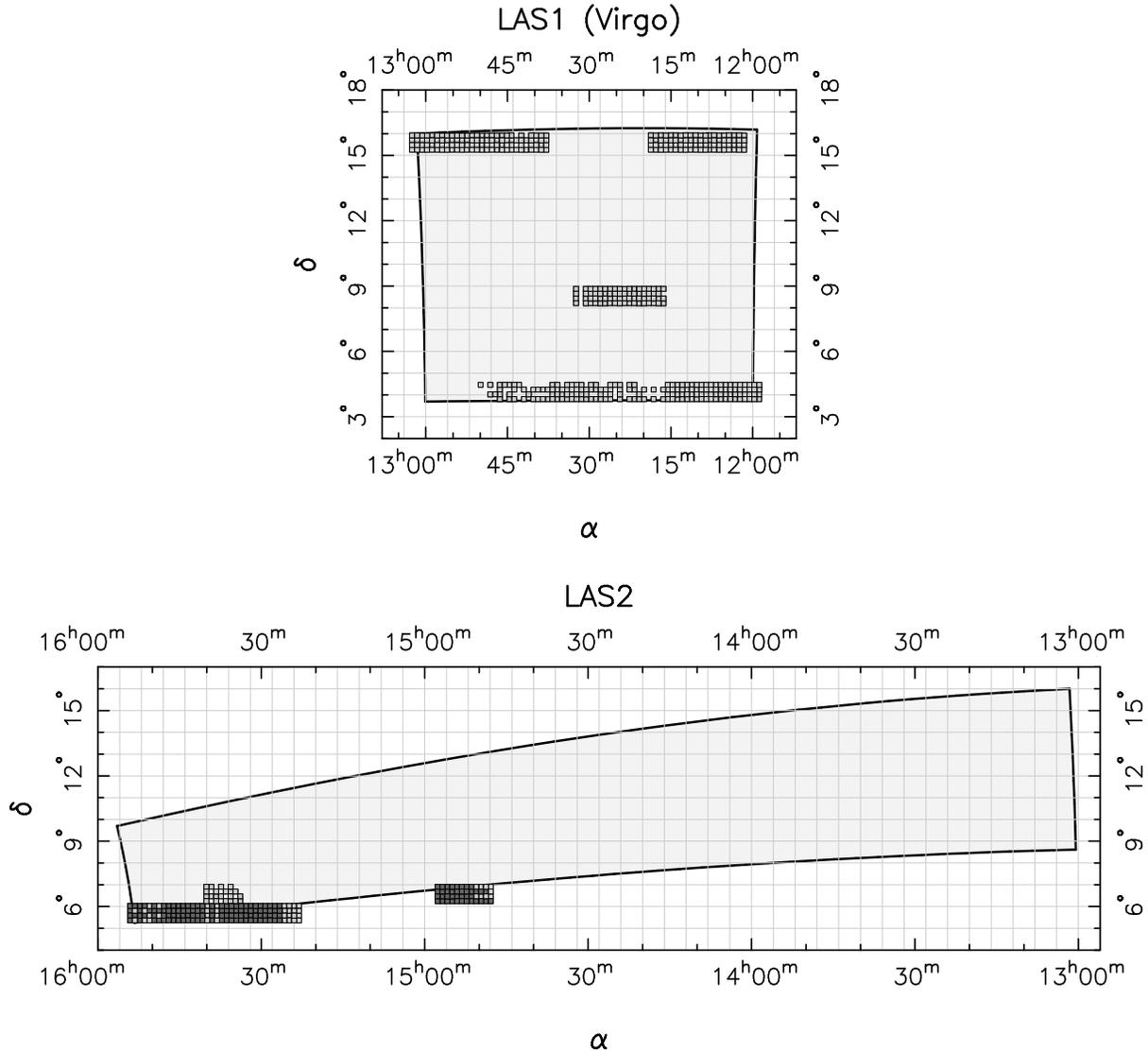}
\hfill}
\caption{LAS data coverage in the EDR database (all of $YJHK$; 
dark grey tiles only) and the EDR+ database (all tiles), for
projects LAS1 and LAS2. Each small square is a detector frame.}
\label{las_coverage}
\end{figure*}

\begin{figure*}
\epsfxsize=163mm
{\hfill
\epsfbox{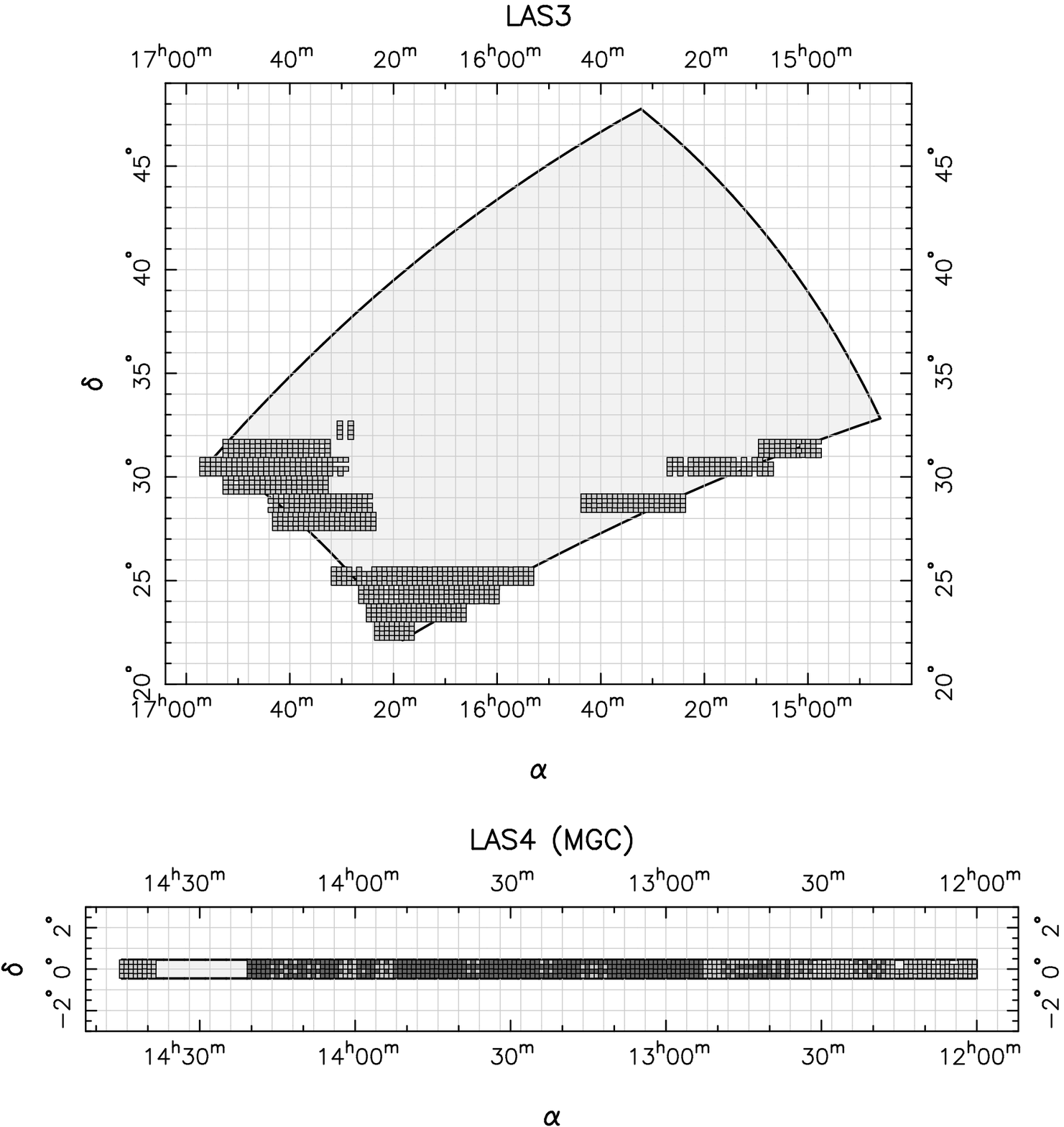}
\hfill}
\caption{LAS data coverage in the EDR database (all of $YJHK$; 
dark grey tiles only) and the EDR+ database (all tiles), for
projects LAS3 and LAS4. LAS3 is currently only observed in $J$. Each
small square is a detector frame.}
\label{las_coverage2}
\end{figure*}

\begin{figure*}
\epsfxsize=157mm
{\hfill
\epsfbox{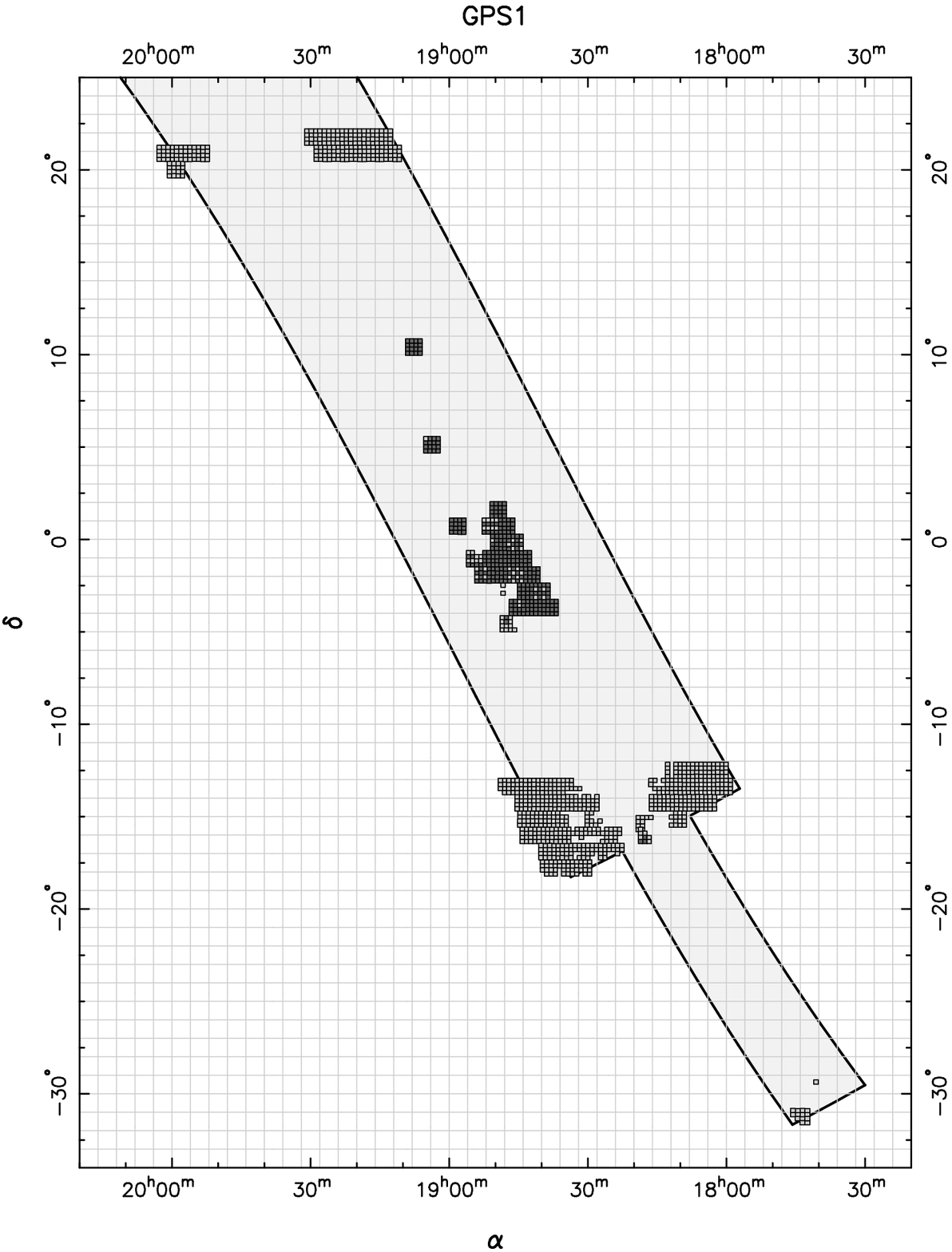}
\hfill}
\caption{GPS coverage in the EDR database (all of $JHK$; dark grey
 tiles) and the EDR+ database (all tiles). EDR database coverage is
 specifically limited to $|{\rm b}|<1^{\circ}$, $-2^{\circ}<{\rm
  l}<107^{\circ}$, and $|{\rm b}|<5^{\circ}$, $30^{\circ}<{\rm
  l}<45^{\circ}$. Areas outside this are $K$-band only, first-epoch
  observations. Each small square is a detector frame.}
\label{gps_coverage}
\end{figure*}

\begin{figure*}
\epsfxsize=149mm
{\hfill
\epsfbox{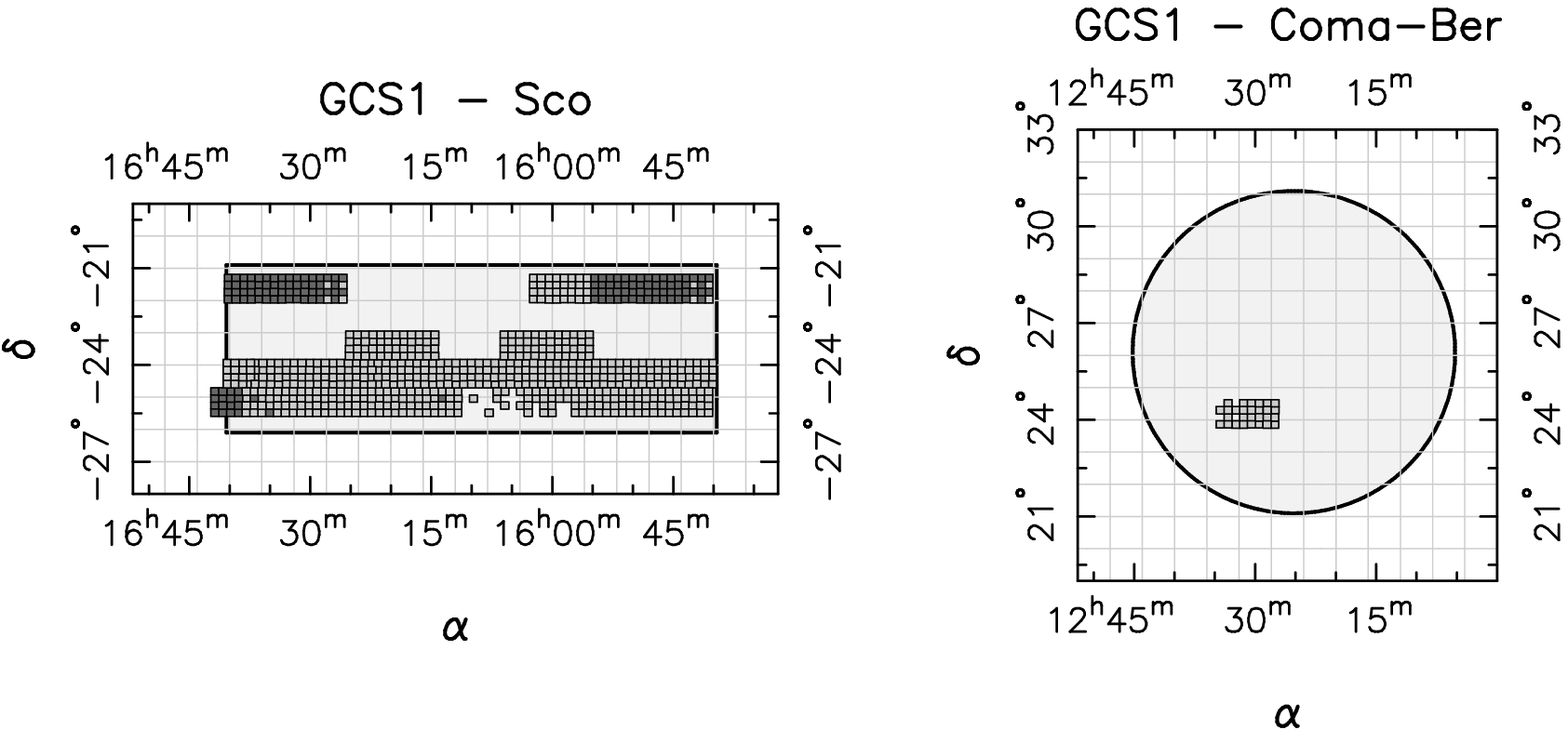}
\hfill}
\caption{GCS coverage in the EDR database (all of $ZYJHK$; dark 
grey tiles) and the EDR+ database (all tiles). Each small
square is a detector frame.}
\label{gcs_coverage}
\end{figure*}

Figures \ref{las_coverage} to \ref{gcs_coverage} provide maps of the sky
coverage for the three surveys. In each map, every detector is
shown. Dark grey tiles mark EDR database fields and light grey tiles
mark the additional fields in the EDR+ database. For the purposes of
comparison, all the coverage plots for the shallow surveys have the
same scale. We now briefly describe the maps for each survey.

In the LAS, data were taken in the projects LAS$1-4$, shown in
Figure \ref{ukidss_survey_areas}. Maps of the areas covered are provided
in Figures \ref{las_coverage} and \ref{las_coverage2}. All four projects
provide data to the EDR+ database, but data in the EDR database is
confined to projects LAS2 and LAS4. The majority of the data in the
EDR database is in LAS4, which covers the eastern half of the
Millennium Galaxy Catalogue (MGC) strip \citep{liske03}. The project
LAS3 contains exclusively first-epoch data in the $J$ band.

GPS data in the EDR+ database are limited to the range of galactic
longitude $-2^{\circ} < {\rm l}< 107^{\circ}$. The coverage map is
provided in Figure \ref{gps_coverage}. Within this region, the EDR
database coverage is specifically limited to $|{\rm b}|<1^{\circ}$,
$-2^{\circ}<{\rm l}<107^{\circ}$ and $|{\rm b}|<5^{\circ}$,
$30^{\circ}<{\rm l}<45^{\circ}$. Care should be taken in interpreting
the GPS catalogues in regions of nebulosity, which can cause
difficulties for the background-following algorithm.

Finally, for the GCS, the EDR+ database holds data for the Sco. star
forming association and a small region in the Coma-Ber open cluster.
Coverage in the EDR database is confined to the Sco. star forming
association. The coverage map is provided in Figure
\ref{gcs_coverage}.

\subsection{Deep survey data}
\label{sec_deep_contents}
  
DXS data span semesters 05A and 05B between the dates 9th April 2005
and 27th September 2005. Data were taken in all four DXS fields. UDS
data are from semester 05B only, from the 12th to the 27th September
2005.

The basic unit of the deep surveys is the Leavstack frame that results
from the implementation schemes detailed in Table
\ref{tab_obs_design}.  For the DXS the frames are interleaved at the
scale $0.2\arcsec$/pix, and are the result of total integration times
of either 500s (05A) or 640s (05B). For the UDS, the frames are
interleaved at the scale $0.133\arcsec$/pix and are the result of a
total integration time of 810s. Leavstack frames are created by
interleaving on a sub-pixel grid and averaging. Interleaving is viable
because the offsets used are small, $<10\arcsec$.

Depth is built up by averaging several Leavstack frames in the same
field. The result is referred to in the WSA as a `Deepleavstack'
frame.  To minimise systematic errors, Leavstack frames are offset
from each other by larger offsets, of order $30\arcsec$. On this
scale, the variation of the pixel scale across the field of view
becomes relevant, and it is necessary to re-sample the Leavstack frames
for accurate registration before averaging.

For both the DXS and UDS, catalogues of sources are produced for all
Leavstack and Deepleavstack frames, and are added to the respective
detection tables, i.e. there can be multiple detections of a
particular source in the same filter. The source tables, however, are
limited to objects detected in Deepleavstack frames only. For the DXS,
the EDR database contains data for fields observed in both $J$ and
$K$, while the EDR+ database additionally contains data where
observations in only one filter have been completed. Since the UDS
comprises a single Tile, querying the EDR and EDR+ databases for the
UDS produces the same result.
  
\subsubsection{DXS coverage and depth}

\begin{figure*}
\epsfysize=150mm
{\hfill
\epsfbox{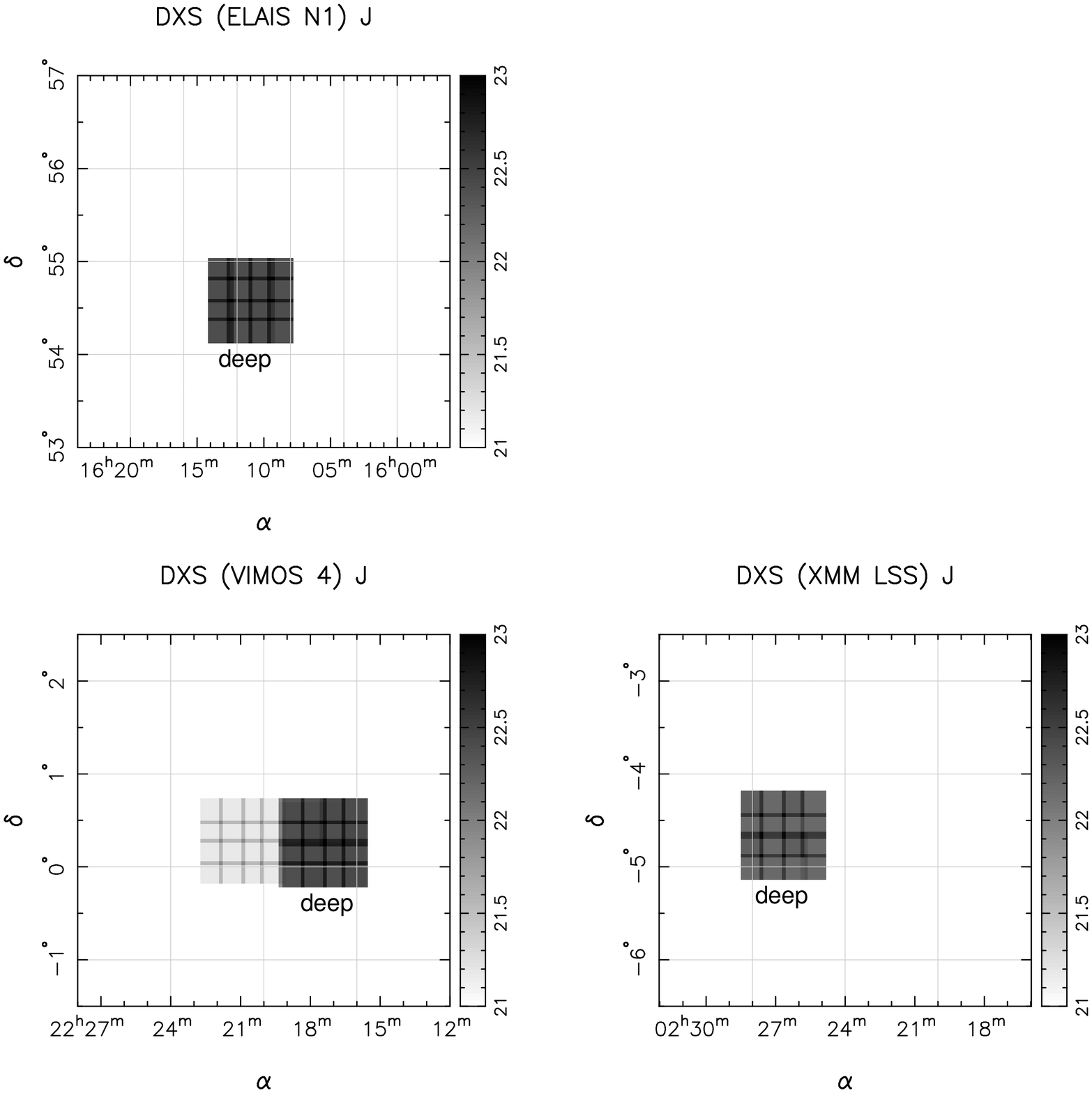}
\hfill}
\caption{DXS coverage and depth in $J$. Tiles that are available from
the WSA as Deepleavstacks are labelled. Grey-shaded key gives depth in $J$.}
\label{dxs_coverage_j}
\end{figure*}

\begin{figure*}
\epsfysize=150mm
{\hfill
\epsfbox{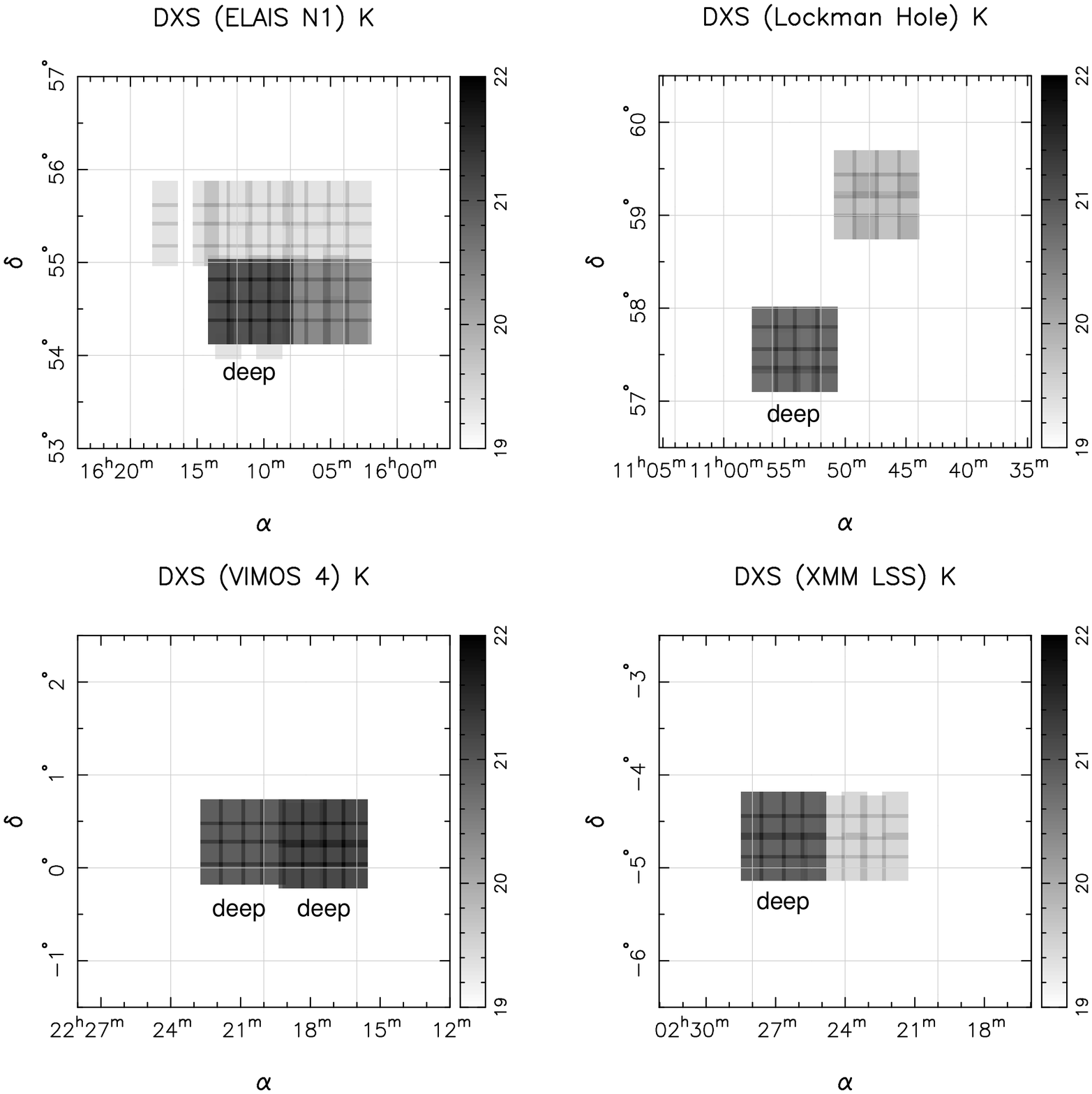}
\hfill}
\caption{DXS coverage and depth in $K$. Tiles that are available from
the WSA as Deepleavstacks are labelled. Grey-shaded key gives depth in $K$.}
\label{dxs_coverage_k}
\end{figure*}

\begin{figure}
\epsfxsize=80mm
{\hfill
\epsfbox{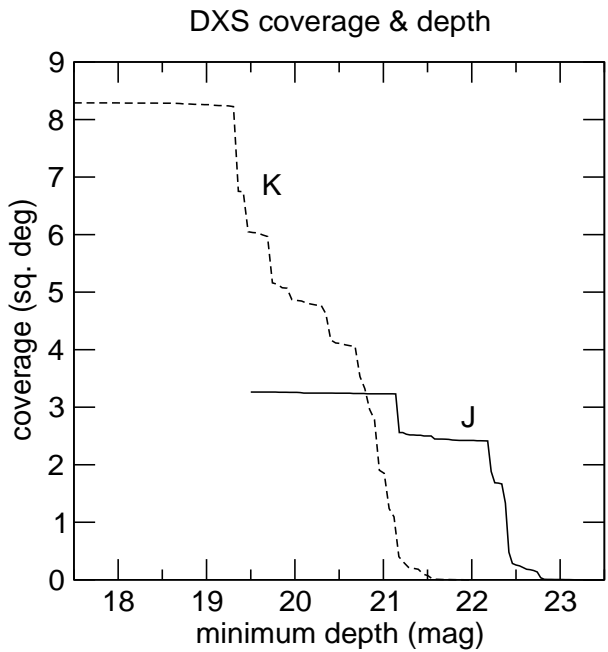}
\hfill}
\caption{DXS areal coverage attaining a given minimum depth as a
function of that minimum depth, for $J$ and $K$.}
\label{dxs_cum_pix_depth}
\end{figure}

The coverage of the DXS fields in $J$ and $K$ is illustrated in
Figures \ref{dxs_coverage_j} and \ref{dxs_coverage_k},
respectively. The grey-scale represents depth, established from
$t_{\rm tot}$, the total integration time at any coordinate, and
scaled by $1.25\mathrm{log_{10}}(t_{\rm tot})$. The total areal
coverage is 3.3 deg$^2$ in $J$ and 8.3 deg$^2$ in $K$. Figure
\ref{dxs_cum_pix_depth} also gives a graphical representation of the
DXS coverage by plotting the area that attains a given minimum depth
as a function of that minimum depth. Deepleavstack frames have been
created only in the fields with the largest total integration
times. The stacking algorithm is similar to that used for creating the
Leavstack frames. The fields are the 3 tiles (2.3 deg$^2$) in $J$ and
5 tiles (3.9 deg$^2$) in $K$ marked `deep' in the coverage
maps. Details of the 32 DXS Deepleavstack multiframes are provided in
Table \ref{tab_dxs_deep_stacks}. The first column provides the field
name, and the second is a code identifying the sub-field, in the form
{\em Tile.XY}, where {\em XY} are coordinates specifying one of four
multiframe positions that make up a Tile. The remaining columns give
coordinates, filter, total integration time $t_{\rm tot}$, and
$5\sigma$ depth.

\begin{table*}
\centering
\begin{tabular}{|l|c|c|c|c|c|c|} 
\hline
Field &  Sub-field & RA (J2000) & Dec(J2000) & Filter & $t_{\rm tot}$ (s) & Depth (mag)\\ 
\hline
ELAIS N1$^{(\dag)}$ & 1.00 &  242.5994000 & 54.5031333 & $J$ & 2640. & 21.42 \\
ELAIS N1$^{(\dag)}$ & 1.10 &  242.5994000 & 54.7233250 & $J$ & 4420. & 21.64 \\
ELAIS N1$^{(\dag)}$ & 1.01 &  242.9817370 & 54.5031333 & $J$ & 6060. & 21.69 \\
ELAIS N1$^{(\dag)}$ & 1.11 &  242.9817370 & 54.7233250 & $J$ & 5060. & 21.62 \\
ELAIS N1 & 1.00 &  242.5994000 & 54.5031333 & $K$ & 4500. & 20.43 \\
ELAIS N1 & 1.10 &  242.5994000 & 54.7233250 & $K$ & 4500. & 20.51 \\
ELAIS N1 & 1.01 &  242.9817370 & 54.5031333 & $K$ & 5000. & 20.53 \\
ELAIS N1 & 1.11 &  242.9817370 & 54.7233250 & $K$ & 3500. & 20.33 \\
Lockman Hole & 1.00 &  163.3623000 & 57.4753556 & $K$ & 3500. & 20.35 \\
Lockman Hole & 1.10 &  163.3623000 & 57.6955472 & $K$ & 2500. & 20.19 \\
Lockman Hole & 1.01 &  163.7756167 & 57.4753556 & $K$ & 2000. & 20.19 \\
Lockman Hole & 1.11 &  163.7756167 & 57.6955472 & $K$ & 2000. & 20.02 \\
XMM-LSS & 1.00 &  36.5752020 & -4.7496444 & $J$ & 5760. & 22.26 \\
XMM-LSS & 1.10 &  36.5803542 & -4.5294528 & $J$ & 6400. & 22.28 \\
XMM-LSS & 1.01 &  36.8012000 & -4.7496444 & $J$ & 5760. & 22.24 \\
XMM-LSS & 1.11 &  36.8012000 & -4.5294528 & $J$ & 5760. & 22.27 \\
XMM-LSS & 1.00 &  36.5752020 & -4.7496444 & $K$ & 6400. & 20.83 \\
XMM-LSS & 1.10 &  36.5803542 & -4.5294528 & $K$ & 6400. & 20.87 \\
XMM-LSS & 1.01 &  36.8012000 & -4.7496444 & $K$ & 6400. & 20.82 \\
XMM-LSS & 1.11 &  36.8012000 & -4.5294528 & $K$ & 3840. & 20.62 \\
VIMOS 4 & 1.00 &  334.2667458 & 0.1698000 & $J$ & 8320. & 22.21 \\
VIMOS 4 & 1.10 &  334.2667458 & 0.3899917 & $J$ & 10880. & 22.38 \\
VIMOS 4 & 1.01 &  334.4869458 & 0.1698000 & $J$ & 9600. & 22.36 \\
VIMOS 4 & 1.11 &  334.4869458 & 0.3899917 & $J$ & 5760. & 22.18 \\
VIMOS 4 & 1.00 &  334.2667458 & 0.1698000 & $K$ & 9120. & 20.82 \\
VIMOS 4 & 1.10 &  334.2667458 & 0.3899917 & $K$ & 12180. & 21.00 \\
VIMOS 4 & 1.01 &  334.4869458 & 0.1698000 & $K$ & 10900. & 20.95 \\
VIMOS 4 & 1.11 &  334.4869458 & 0.3899917 & $K$ & 9620. & 20.97 \\
VIMOS 4 & 2.00 &  335.1420250 & 0.1817444 & $K$ & 7680. & 20.90 \\
VIMOS 4 & 2.10 &  335.1420250 & 0.4019361 & $K$ & 8320. & 20.91 \\
VIMOS 4 & 2.01 &  335.3622250 & 0.1817444 & $K$ & 8960. & 20.92 \\
VIMOS 4 & 2.11 &  335.3622250 & 0.4019361 & $K$ & 8960. & 20.88 \\
\hline  
\end{tabular}
\caption{DXS Deepleavstack multiframes. The field, sub-field code,
base coordinates, filter, total Integration time in seconds and
5$\sigma$ point source sensitivity of each multiframe are
listed. $^{(\dag)}$These four $J$ stacks are affected by the first sky
noise problem discussed in the text.}
\label{tab_dxs_deep_stacks}
\end{table*}

\begin{flushleft}
{\em Caveats}
\end{flushleft}

At the present time, we are aware of two problems affecting the DXS
Leavstack and Deepleavstack frames, both related to the treatment of
the noise in the sky.

The stacking algorithm that produces the Leavstack frames has a
default minimum value of the sky noise of 2.5 counts, which is a
safety feature. In a small number of $J$ band Leavstack frames the
true value is lower. The consequence is that these frames are
incorrectly weighted in forming the Deepleavstack frame, so the
combination is not optimal, and the frame is not as deep as it should
be. The Deepleavstack frames affected are the first four listed in
Table \ref{tab_uds_deep_stacks}.

The second problem is that the stacking and object detection
algorithms operate in integer format. This has the consequence that in
the deepest $J$ band Deepleavstack frames, the sky noise is
under-sampled, meaning that there is a digitisation noise
contribution. The result is that the sky noise is greater than it need
be, and so the detection limits are not as deep as they should be.

Both problems will be remedied in DR1.

\begin{table*}
\centering
\begin{tabular}{|l|c|c|c|c|c|} 
\hline
Quadrant & RA (J2000) & Dec(J2000) & Filter & $t_{\rm tot}$ (s) & Depth \\ 
\hline
North East & 34.7149375 & -4.7893583 & $J$ & 3160.  & 22.47 \\
North West & 34.2424791 & -4.7893583 & $J$ & 3160.  & 22.28 \\
South East & 34.7151208 & -5.2923027 & $J$ & 3160.  & 22.25 \\
South West & 34.2422958 & -5.2923027 & $J$ & 3160.  & 22.26 \\
North East & 34.7149375 & -4.7893583 & $K$ & 5240.  & 21.11 \\
North West & 34.2424791 & -4.7893583 & $K$ & 5240.  & 21.01 \\
South East & 34.7151208 & -5.2923027 & $K$ & 5240.  & 21.00 \\
South West & 34.2422958 & -5.2923027 & $K$ & 5240.  & 21.10 \\
\hline  
\end{tabular}
\caption{UDS deep stacks. The centre coordinates, filter,
total Integration time in seconds and 5$\sigma$ point source
sensitivity of each quadrant are listed.}
\label{tab_uds_deep_stacks}
\end{table*}

\subsubsection{UDS coverage and depth}

The UDS covers a single Tile and so is not illustrated. The total
integration times and depths reached in $J$ and $K$ are summarised in
Table \ref{tab_uds_deep_stacks}, where the area covered has been
divided into quadrants (as explained below).

Instead of using the WFCAM pipeline software, the UDS Leavstacks were
combined using TERAPIX's SWarp package \citep{bertin02}, with a
lanczos3 kernel. The UDS Leavstacks were first mosaiced together to
make single whole-Tile frames.  Since the UDS data are Microstepped to
1/3 of a WFCAM pixel, the final $J$ and $K$ interleaved and stacked
Tiles are $\sim 24000 \times 24000$ pixels, occupying some 3 Gb each
on disk. To ease handling, these have been divided into four
quarters. In the WSA, these quarters are referred to as
`Mosaicdeepleavstack' frames.  Unlike all other stacked data in the
WSA, astrometry of the deep UDS stacks uses tangential projection
(TAN) rather than the zenith polynomial projection (ZPN) that is
standard for WFCAM.

\begin{flushleft}
{\em Caveats}
\end{flushleft}

The source detection algorithm used to generate object catalogues is
not optimal for the oversampled data generated by the $3 \times 3$ UDS
Microstepping strategy. The result is that bright sources tend to be
split into multiple sources around the edges. This problem will be
rectified in DR1.

\section{The WFCAM Science Archive}
\label{sec_wsa}

In this section we outline the architecture of the WSA in its current
state, and describe the principal means by which users can access the
data. The WSA will continue to develop over the course of the 2-year
plan. Upgrades of the WSA will be announced in future data releases as
they occur. We refer the reader to \citet{hambly06} for more
information.

\subsection{WSA architecture}
\label{sec_wsa_arch}

Pipeline-processed images and catalogues are held in the WSA, hosted
in Edinburgh by WFAU. Image data are stored in the same MEF format
produced by the pipeline, with a primary header and one header per
extension. The MEF files are the {\em multiframes} referred to
previously. The object data from the individual catalogues are
extracted upon ingest and placed in a relational database running on
Microsoft SQL Server 2000. The individual catalogue files are
nevertheless retained, and are stored in MEF format.

The multiframe is the principal component of the WSA. The most common
multiframe corresponds to a single WFCAM Integration. Multiframes
exist for all observation types (e.g. dark, flat, object) and data
products (e.g. Normal, Leav \& Stack). The WSA refers to the primary
header of a Multiframe as extension 1 (via the attribute {\tt extNum}
-- see below). Extensions 2 to 5 correspond to the headers and pixel
data for detectors 1 to 4 respectively. The primary header includes
{\em metadata} applicable to all objects in the multiframe, such as
the airmass and filter.  The subsidiary headers include metadata
applicable to the individual detectors, such as the average
ellipticity for point sources and the astrometry coefficients.  Each
multiframe is assigned a unique ID number, {\tt multiframeID}, which
is an attribute in many tables held in the archive.  The attribute
{\tt multiframeID} is used, often in conjunction with the attribute
{\tt extNum}, to link the multiframe records held across tables (see
Section \ref{sec_wsa_access}).

All tables and their attributes are listed and described in the WSA
`schema browser'. The multiframe metadata are provided in the database
in entries in three main tables. The metadata from the primary headers
are held in the {\tt Multiframe} table. Each multiframe in the archive
is represented by a single row in the {\tt Multiframe} table, with
columns for attributes. The table {\tt MultiframeDetector} contains
metadata applicable to the individual detector frames of a
multiframe. This excludes the astrometry calibration coefficients,
which are held in the third main table, {\tt CurrentAstrometry}. A
multiframe with four extensions (detectors) gives rise to four
corresponding entries (rows) in {\tt MultiframeDetector} and {\tt
CurrentAstrometry}.

Every detection in every stack multiframe is recorded in a row in the
{\em detection table} for the particular survey, e.g. {\tt
gpsDetection}. This means that there can be several rows for the same
object, corresponding to detections in different filters, or multiple
observations in the same filter. All survey detection tables contain
the same attributes. A list of the attributes is provided in Table
\ref{tab_detection_table}. The {\em source table} for each survey
(e.g. {\tt gpsSource}) summarises the data in the corresponding
detection table by a single row for every object, listing details of
every detection of that object. Matching of multiple detections is
achieved on the basis of positional coincidence. The details of the
source matching algorithm are provided in \citet{hambly06}. The number
of attributes recorded in the source table, for each detection, is a
subset of the attributes recorded in the detection table, to keep the
number of columns in the source table manageable. The same subset is
used for detections in the source tables of all the surveys. However,
the number of possible detections of an object depends on the filter
coverage, and number of repeat passes (Table \ref{tab_survey_final}),
and so is different for each survey. (Note that photometry in the same
filter, at different epochs, is distinguished by e.g. {\tt j\_1}, {\tt
j\_2}, for the two $J$ epochs of the LAS.) Table
\ref{tab_source_table} provides the list of attributes in the table
{\tt lasSource}. A log of the source merging is recorded in a table
for each survey e.g. {\tt lasMergeLog}.  The attributes {\tt
multiframeID}, {\tt frameSetID} and {\tt SeqNum} are provided in the
source tables to allow cross-referencing. In this way, for a
particular object in a source table, it is possible to recover the
full set of attributes in the detection table, for each
detection. Examples of how to cross reference are provided in Section
\ref{sec_wsa_access}.

\subsection{WSA access}
\label{sec_wsa_access}

The WSA is accessed at the URL {\tt http://surveys.roe.ac.uk/wsa}.
For a period of eighteen months from the release date, access to
UKIDSS data is restricted to astronomers residing in ESO-member
countries. For the EDR this release date was 10th February 2006. Once
this restricted period has expired, the data are freely available
worldwide. Access to UKIDSS data during the restricted period requires
registering for an account, organised through ``community contacts" at
local institutions. The registration process is described on the
UKIDSS web pages.\footnote{http://www.ukidss.org/archive/archive.html}

\subsubsection{Access to pixel data}
\label{sec_wsa_pix_access}

Pixel data are obtained from the WSA using menu driven interfaces.
Users can obtain details of all multiframes within an area of sky
through the `Archive Listing' interface, selecting on database (EDR or
EDR+), survey, observation type (e.g. dark, flat, object), frame type
(e.g. normal, leav, stack), filter and observation date. The result
of a query is a table of multiframes that satisfy these criteria, with
their attributes. Links are provided for each multiframe to a jpeg
preview image, to the (compressed) MEF image file itself, and to the
catalogue of detections for that multiframe (in the case of stack
frames). Alternatively the same list of multiframes could have been
obtained with an appropriate free-form SQL query (explained below), and
the MEF image file obtained by entering the {\tt multiframeID} in the
Archive Listing menu.

Instead of searching the archive by multiframe, it is possible to
extract cut-out images, of a requested size, around a given position
using the `GetImage' and `MultiGetImage' interfaces. Similar to the
Archive Listing interface, the GetImage interface enables users to
refine the search by database, survey, observation type, frame type,
and filter, or to provide the {\tt multiframeID}. At present, the
extraction region is confined by the boundaries of the detector frame
in which the given position is located and therefore limited in size
by the detector dimensions. However, in the future, mosaicing of
adjacent multiframes will be implemented and it will be possible to
extract regions as large as $90\arcmin\times90\arcmin$ in size. The
MultiGetImage interface accepts a list of RA and Dec coordinates,
currently limited to a maximum of 500. Once a submitted MultiGetImage
request is completed, an email is sent to the user containing a link
to their results. Jpeg preview images and FITS image files are
produced and presented in web page tables and a PDF file. A tar save-set
of the FITS files is also generated.

\subsubsection{Access to catalogue data}
\label{sec_wsa_cat_access}

The power of the WSA lies in the possibility of flexibly querying the
detection and source tables. Three methods of access are offered (not
counting the ability to download individual multiframe detection
catalogues, explained in Section \ref{sec_wsa_pix_access}). The first
option is a simple `Region' search where the user selects a table
(detection or source) by survey. The application returns all objects
in the table within the chosen radius (currently limited to $\le
90\arcmin$) of the position provided. A more sophisticated search may
be carried out using the `Menu query' interface, which the user
completes in stages. After selecting the relevant table, as above, the
user is presented with a list of table attributes, choosing the ones
to output. One then has the option of delimiting on any of the
selected attributes. From here, the request may be submitted or,
alternatively, the underlying SQL query generated by the menu
interface may be edited directly. In the latter case, the SQL text is
passed to the free-form SQL query interface for amendment.

The final option is the free-form SQL interface. This option is the
most powerful, allowing users to fully exploit the archive, including
the possibility of combining tables. The WSA web pages provide a
cook-book for users unfamiliar with SQL. Below we provide some simple
worked examples to illustrate the possibilities.

\begin{flushleft}
{\em Example 1: Return coordinates and $2\arcsec$-diameter aperture
  $YJHK$ photometry of objects in a 1 deg$^2$ rectangular region in
  the LAS}
\end{flushleft}
This is a simple search that could be undertaken with the Menu query
interface described above, which illustrates the fundamentals of SQL
queries. The fact that multi-filter information is needed dictates
that the table {\tt lasSource} is required. The query is achieved with
the following command (note that WSA SQL is case insensitive,
capitalisation here is for clarity only):
\begin{flushleft}
{\tt
SELECT ra,dec,yAperMag3,j\_1AperMag3, \\
\hspace{3mm} hAperMag3,kAperMag3 \\
FROM lasSource\\
WHERE ra BETWEEN 190 AND 191\\
\hspace{3mm} AND dec >= -0.5 AND dec <= 0.5
}
\end{flushleft}
All SQL queries start with the {\tt SELECT} statement followed by a
list of required attributes, which in this case are the coordinates
and appropriate aperture magnitudes. The second part of a simple SQL
select is the {\tt FROM} statement followed by the table name. The
{\tt WHERE} clause in the example is optional and is used here to
refine the search to a $1^{\circ} \times 1^{\circ}$ region. Note that
both RA and Dec are expressed in decimal degrees and that the SQL `{\tt
ra BETWEEN 190 AND 191}' is identical to `{\tt ra >= 190 AND ra
<=191}'. In addition, note that this example returns duplicated objects
that lie in detector overlap regions. To eliminate these, another line
of SQL must be appended to select only the primary entry using
the {\tt priOrSec} attribute: {\tt AND (priOrSec=0 OR priOrSec=frameSetID)}

As an example of the flexibility of the WSA, one could use a simple
command of this type to select rare objects of unusual colour, and use
the output to check the reality of the candidates using
MultiGetImage. A variety of different formats for the output of a SQL
query is offered. By outputting a list of RA and Dec of objects
satisfying certain colour criteria, and selecting uncompressed ASCII
output, the resulting file may be edited to remove the header, and
then directly read by the MultiGetImage interface, to return FITS
images of each candidate.

\begin{flushleft}
{\em Example 2: Find all $K$-band detector frames in the GPS with fewer
than 30000 detected objects}
\end{flushleft}
To achieve this, the attributes {\tt filterID} and {\tt tableRows} in
the table {\tt MultiframeDetector} provide the filter name and the
number of objects detected.\footnote{{\tt filterID}=1 to 8 corresponds
  to filters $Z$, $Y$, $J$, $H$, $K$, $H_2$, $Br\gamma$, and dark
  respectively. The SQL `{\tt SELECT * FROM Filter}' gives further
  details.}  However, there is no attribute in {\tt MultiframeDetector} which
can be used to identify the survey. A SQL `join' with table {\tt
  Multiframe} must be made, using the attribute {\tt multiframeID} as
the common link, in order to access the attribute {\tt project}
(a survey is divided into several projects, for administrative
purposes).  The following SQL provides the required result:
\begin{flushleft}
{\tt
SELECT M.multiframeID,extNum \\
FROM MultiframeDetector AS MD, Multiframe AS M \\
WHERE tableRows < 30000 \\
\hspace{3mm} AND M.multiframeID = MD.multiframeID \\
\hspace{3mm} AND project LIKE "U/UKIDSS/GPS\%" \\
\hspace{3mm} AND M.filterID=5 \\
\hspace{3mm} AND frameType LIKE "\%stack" \\
}
\end{flushleft}
Here, both table names are relabelled with the {\tt AS} function to
simplify the SQL. To prevent ambiguities, attributes common to both
tables, such as {\tt filterID}, must be uniquely specified by
prefixing them with the relevant table name, or as in the example
above, its relabelled name, e.g. {\tt M.filterID}. The join is
achieved by stipulating {\tt M.multiframeID = MD.multiframeID}.  The
{\tt \%} sign is a wildcard, used in conjunction with {\tt LIKE}, to
include all GPS projects, and all types of stack frame (including
Stack and Leavstack).

\begin{flushleft}
{\em Example 3: Return the distribution of depths reached
in GCS $J$-band stacked detector frames}
\end{flushleft}
In this example, we use the definition of point-source $5\sigma$ depth
given by equation (\ref{eq_depth}), explained in Section
\ref{sec_qc}. This formula requires the attributes {\tt photZPCat},
{\tt skyNoise} and {\tt AperCor3}, respectively the photometric zero
point $m_0$, the background noise $\sigma_{\rm sky}$, and the aperture
correction $m_{ap}$, from the table {\tt MultiframeDetector}, as well
as the attribute {\tt expTime}, the exposure time $t_{\rm exp}$, from
the table {\tt Multiframe}. The formula also requires the calculation
of $N$ the number of pixels in the aperture, which depends on the
pixel size, given by ${\tt xPixSize}$ in the table {\tt
CurrentAstrometry} (recall that the pixel size is smaller in an
interlaced frame). Then the number of pixels is given by $N=\pi/${\tt
xPixSize}$^2$, for the $2\arcsec$-diameter aperture used.

Since attributes are used from the tables {\tt MultiframeDetector},
{\tt Multiframe}, and {\tt CurrentAstrometry}, a three-way join is
needed:
\begin{flushleft}
{\tt
SELECT CAST(ROUND(T.depth*10.0,0) AS INT)/10.0 \\
\hspace{2mm} AS depthBin, COUNT(*) \\
FROM (\\
\hspace{8mm}  SELECT photZPCat \\
\hspace{12mm} -2.5*LOG10(5.0*skyNoise* \\
\hspace{12mm} SQRT(1.2*3.141593)/(xPixSize*expTime))\\
\hspace{12mm}  -AperCor3 AS depth \\
\hspace{8mm}  FROM   MultiframeDetector AS MFD,  \\
\hspace{12mm}	     Multiframe AS M, \\
\hspace{12mm}	     CurrentAstrometry AS CA  \\
\hspace{8mm}   WHERE M.multiframeID=MFD.multiframeID  \\
\hspace{12mm}  AND   CA.MultiframeID=MFD.MultiframeID  \\
\hspace{12mm}	     AND CA.extNum=MFD.extNum \\
\hspace{12mm}        AND project LIKE "u/ukidss/gcs\%"\\
\hspace{12mm}	     AND M.filterID=3  \\
\hspace{12mm}	     AND frameType LIKE "\%stack" \\
\hspace{8mm}   ) AS T \\
GROUP BY CAST(ROUND(T.depth*10.0,0) AS INT)/10.0 \\
ORDER BY depthBin
}
\end{flushleft}
This is an example of a nested select. The inner select statement
returns a table, labelled {\tt T} here, which is the list of depths
required. The table {\tt T} is then subject to the external select
statement.  To return a distribution of depths, the above SQL does
three things: 1) rounds depths to the nearest 0.1 mag with the function
{\tt CAST(ROUND(T.depth*10.0,0) AS INT)/10.0}, 2) groups these binned
depths together with the {\tt GROUP BY} statement, 3) counts the
number of binned depths in each group with {\tt COUNT(*)}.  The {\tt
ORDER BY} statement simply returns results in order of ascending bin
magnitude.

For further examples and more information on access to the WSA, we
refer the reader to the WSA Cookbook, Schema Browser and Data
Access/Overview via the web address given at the beginning of this
section \citep[see also][]{hambly06}.

\section{Summary}

The UKIDSS EDR is the first ESO-wide release of UKIDSS data.  The data
constitute $\sim 1\%$ of the whole of UKIDSS, and are are of similar
quality to, or marginally worse than, the expected data quality in
future releases.  Although this is only a small fraction of what
UKIDSS will deliver at the end of its 7-year plan, the EDR is already
comparable to the size of 2MASS in terms of the number of photons
collected.

The EDR includes data from each of the five surveys that make up
UKIDSS. The smaller EDR database contains $\sim 50$ deg$^2$ of pixel
and catalogue data in the regions where the full complement of filters
for the particular survey is complete. In addition we have released an
extended database, the EDR+, that contains all data in the EDR
database plus extra data that have passed QC, but in regions where the
full complement of filters is incomplete. The EDR+ database
constitutes approximately 220 deg$^2$.

In this paper, we have described the full data-train, from acquisition
to release. We have given some background information to UKIDSS and an
overview of the goals of the surveys. We have provided details of the
observational implementation, outlined our quality control and
calibration procedures, quantified the quality of the image and
catalogue data and finally given instructions and worked examples for
access to the WSA.

The first large release of UKIDSS data to the ESO community, DR1, will
occur in mid-2006. DR1 will contain all data observed in 05A and
05B, that pass QC. We refer the reader to the UKIDSS website ({\em
http://www.ukidss.org}) for news and progress updates.


\vspace{5mm}
\begin{flushleft}
{\bf Acknowledgements}
\end{flushleft}

SD thanks the U.K. Particle Physics and Astronomy Research Council for
6 months funding as Consortium Science Verifier.  We are grateful to
Chris Wolf for discussions on quality control.

\appendix

\section{Tables of contents of the WSA}

For orientation, we tabulate the attributes found in all WSA detection
tables (Table \ref{tab_detection_table}) and, as an example since they
are survey specific, the attributes in the {\tt lasSource} table
(Table \ref{tab_source_table}). For more complete and up-to-date
information, users should refer to the online versions of these tables
found in the WSA schema browser under the web address {\em
http://surveys.roe.ac.uk/wsa}.

\begin{table*}
\centering
\begin{tabular}{@{}lll@{}}
\hline
Name		&Unit	&Description	\\
\hline
{\tt multiframeID}	& 	&Unique UD (UID) of the relevant multiframe	\\
{\tt extNum}	        & 	&Extension number of the multiframe	\\
{\tt cuEventID}	& 	&UID of curation event giving rise to this record \\
{\tt seqNum}          &	&Running number of this detection \\
{\tt filterID}	& 	&UID of combined filter (assigned in WSA) \\
{\tt isoFlux}         &ADU	&Instrumental isophotal flux counts	\\
{\tt isoMag}	        &mag	&Calibrated isophotal magnitude	\\
{\tt x}               &pix	&X coordinate of detection \\
{\tt xErr}            &pix	&Error in X coordinate \\
{\tt y  }             &pix	&Y coordinate of detection \\
{\tt yErr}            &pix	&Error in Y coordinate \\
{\tt gauSig}          &pix	&RMS of axes of ellipse fit \\
{\tt ell}             & 	&1-b/a, where a,b=semi-major,-minor axis \\
{\tt pa}      	&deg    &	Orientation of ellipse fit to x axis \\
{\tt aProf[1-8]}        &pix	&No. pixels above a series of thresholds relative to local sky	\\
{\tt pHeight}         &ADU	&Highest pixel value above sky\\
{\tt pHeightErr}	&ADU	&Error in peak height\\
{\tt aperFlux[1-13]}  &ADU	&Aperture fluxes of diameter (arcsec): 1, $\sqrt{2}$, 2, 2$\sqrt{2}$, 4, 4$\sqrt{2}$, 8, 10, 12, 14, 16, 20, 24\\
{\tt aperFlux[1-13]err}&  ADU	&Error in aperture fluxes [1-13]\\
{\tt aperMag[1-13]}	&mag	&Calibrated aperture corrected aperture magnitudes [1-13]\\
{\tt aperMag[1-13]err} &  mag	&Error in calibrated aperture magnitudes [1-13]\\
{\tt petroRad}	&pix	& Petrosian radius $r_p$, defined in Yasuda et al. 2001 AJ 112 1104\\
{\tt kronRad}	        &pix	&Kron radius $r_k$, defined in Bertin and Arnouts 1996 A\&A Supp 117 393\\
{\tt hallRad}	        &pix	&Hall radius $r_h$,  e.g. Hall \& Mackay 1984 MNRAS 210 979\\
{\tt petroFlux}	&ADU	&Petrosian flux within circular aperture to $2 r_p$ \\
{\tt petroFluxErr}	&ADU	&Error on Petrosian flux\\
{\tt petroMag}	&mag	&Calibrated magnitude from Petrosian flux\\
{\tt petroMagErr}	&mag	&Error on calibrated Petrosian magnitude	\\
{\tt kronFlux}	&ADU	&Kron flux within circular aperture to $2r_k$\\
{\tt kronFluxErr}	&ADU	&Error on Kron flux\\
{\tt kronMag}	        &mag	&Calibrated Kron magnitude from Kron flux\\
{\tt kronMagErr}	&mag	&Error on calibrated Kron magnitude\\
{\tt hallFlux}	&ADU	&Hall flux within circular aperture to $2r_h$: Alternative total flux\\
{\tt hallFluxErr}	&ADU	&Error on Hall flux\\
{\tt hallMag}	        &mag	&Calibrated magnitude from Hall flux\\
{\tt hallMagErr}	&mag	&Calibrated error on Hall magnitude\\
{\tt errBits}	        &	&Processing warning/error bitwise flags\\
{\tt sky}	        &ADU	&Local interpolated sky level from background tracker\\
{\tt skyVar}	        &ADU	&Local estimate of variation in sky level around image\\
{\tt deblend}	        & 	&Flag for parent of deblended deconstruct (redundant; only deblended images kept)\\
{\tt ra}              &deg    &	Celestial Right Ascension\\
{\tt dec}             &deg    &	Celestial Declination\\
{\tt cx	}        &	&Unit vector of spherical coordinate\\
{\tt cy}		&      	&Unit vector of spherical coordinate\\
{\tt cz	}        & 	&Unit vector of spherical coordinate\\
{\tt htmID}	        &	&HTM index (20 digits) for equatorial coordinates\\
{\tt l}	        &deg    &	Galactic longitude\\
{\tt b}	        &deg    &	Galactic latitude\\
{\tt lambda}	        &deg    &	SDSS system spherical coordinate 1\\
{\tt eta}	        &deg    &	SDSS system spherical coordinate 2\\
{\tt class}     	& 	&Flag indicating most probable morphological classification\\
{\tt classStat}  	& 	&N(0,1) stellarness-of-profile statistic\\
{\tt psfFlux}	        &ADU	&PSF-fitted flux ({\bf not currently implemented})\\
{\tt psfFluxErr}	&ADU	&Error on PSF-fitted flux ({\bf not currently implemented})\\
{\tt psfMag}		&mag	&PSF-fitted calibrated magnitude ({\bf not currently implemented})\\
{\tt psfMagErr}	&mag	&Error on PSF-fitted calibrated magnitude ({\bf not currently implemented})\\
{\tt psfFitX}		&pix	&PSF-fitted X coordinate ({\bf not currently implemented})\\
{\tt psfFitXerr}	&pix	&Error on PSF-fitted X coordinate ({\bf not currently implemented})\\
{\tt psfFitY}		&pix	&PSF-fitted Y coordinate ({\bf not currently implemented})\\
{\tt psfFitYerr}	&pix	&Error on PSF-fitted Y coordinate ({\bf not currently implemented})\\
{\tt psfFitChi2}	&	&standard normalised variance of PSF fit ({\bf not currently implemented})\\
{\tt psfFitDof}       &	&No. degrees of freedom of PSF fit ({\bf not currently implemented})\\
\hline
\end{tabular}
\caption{List of parameters in the WSA detection table. Note that at the present time,
PSF fitted fluxes and Sersic fluxes are not implemented. WSA attributes for these
fluxes currently contain default values.}
\label{tab_detection_table}
\end{table*}

\begin{table*}
\centering
\begin{tabular}{@{}lll@{}}
\hline
Name		&Unit	&Description	\\
\hline
{\tt SerFlux1D}       &ADU	&1D Sersic flux ({\bf not currently implemented})\\
{\tt SerMag1D}        &mag	&Calibrated 1D Sersic flux ({\bf not currently implemented})\\
{\tt SerScaleLen1D}   &	&Sersic scale length ({\bf not currently implemented})\\
{\tt SerIdx1D}        & 	&Power law index ({\bf not currently implemented})\\
{\tt SerFit1DChi2}    & 	&Error in 1D fit ({\bf not currently implemented})\\
{\tt SerFitNu1D}      & 	&1D Sersic fit nu ({\bf not currently implemented})\\
{\tt SerFlux2D }      &ADU	&2D Sersic flux ({\bf not currently implemented})\\
{\tt SerMag2D  }      &mag	&Calibrated 2D Sersic flux ({\bf not currently implemented})\\
{\tt SerScaleLen2D}   & 	&Scale length ({\bf not currently implemented})\\
{\tt SerIdx2D  }      & 	&Power law index ({\bf not currently implemented})\\
{\tt SerFit2DChi2}    &   	&Error in 2D fit ({\bf not currently implemented})\\
{\tt SerFitNu2D}      & 	&2D Sersic fit nu ({\bf not currently implemented})\\
{\tt ppErrBits}	&	&Additional WFAU post-processing error bits (currently a place holder)\\
{\tt deprecated}	& 	&Code for a current (=0) or deprecated (!=0) detection \\
{\tt objID}	        & 	&Unique identifier for this detection\\
\hline
\end{tabular}
\contcaption{List of parameters in the WSA detection table. }
\end{table*}

\begin{table*}
\centering
\begin{tabular}{@{}lll@{}}
\hline
Name		&Unit	&Description	\\
\hline
{\tt sourceID}	&	&UID of merged detection\\
{\tt cuEventID}	&	&UID of curation event giving rise to this record\\
{\tt frameSetID}	&	&UID of the set of frames that this merged source comes from\\
{\tt ra}		&Degrees&	Celestial Right Ascension\\
{\tt dec}		&Degrees&	Celestial Declination\\
{\tt sigRa}		&Degrees&	Uncertainty in RA\\
{\tt sigDec}		&Degrees&	Uncertainty in Dec\\
{\tt epoch}		&Years	&Epoch of position measurement\\
{\tt muRa}		&mas/yr	&Proper motion in RA direction ({\bf not currently implemented})\\
{\tt muDec}		&mas/yr	&Proper motion in Dec direction ({\bf not currently implemented})\\
{\tt sigMuRa}		&mas/yr	&Error on proper motion in RA direction ({\bf not currently implemented})\\
{\tt sigMuDec}	&mas/yr	&Error on proper motion in Dec direction ({\bf not currently implemented})\\
{\tt chi2}		&	&Chi-squared value of proper motion solution ({\bf not currently implemented})\\
{\tt nFrames}		&	&No. of frames used for this proper motion measurement ({\bf not currently implemented})\\
{\tt cx}		&	&unit vector of spherical co-ordinate\\
{\tt cy}		&	&unit vector of spherical co-ordinate\\
{\tt cz}		&	&unit vector of spherical co-ordinate\\
{\tt htmID}		&	&HTM index, 20 digits, for co-ordinate\\
{\tt l}		&Degrees&	Galactic longitude\\
{\tt b}		&Degrees&	Galactic latitude\\
{\tt lambda}		&Degrees&	SDSS system spherical co-ordinate 1\\
{\tt eta}		&Degrees&	SDSS system spherical co-ordinate 2\\
{\tt priOrSec}	&	&Seam code for a unique (=0) or duplicated (!=0) source (i.e. overlap duplicates)\\
{\tt ymj\_1Pnt}	&mag	&Point source colour Y-J\\
{\tt ymj\_1PntErr}     &mag	&Error on point source colour Y-J\\
{\tt j\_1mhPnt}	&mag	&Point source colour J-H\\
{\tt j\_1mhPntErr}	&mag	&Error on colour J-H	\\
{\tt hmkPnt}		&mag	&Point source colour H-K\\
{\tt hmkPntErr}	&mag	&Error on point source colour H-K\\
{\tt ymj\_1Ext}	&mag	&Extended source colour Y-J\\
{\tt ymj\_1ExtErr}	&mag	&Error on extended source colour Y-J\\
{\tt j\_1mhExt}	&mag	&Extended source colour J-H\\
{\tt j\_1mhExtErr}	&mag	&Error on extended source colour J-H\\
{\tt hmkExt}		&mag	&Extended source colour H-K\\
{\tt hmkExtErr}	&mag	&Error on extended source colour H-K\\
{\tt mergedClassStat}	& 	&Merged N(0,1) stellarness-of-profile statistic\\
{\tt mergedClass}	&	&Class ($1|0|-1|-2|-3|-9$=galaxy$|$noise$|$stellar$|$probableStar$|$probableGalaxy$|$saturated)\\
{\tt pStar}		&	&Probability that the source is a star\\
{\tt pGalaxy}		&	&Probability that the source is a galaxy\\
{\tt pNoise}		&	&Probability that the source is noise\\
{\tt pSaturated}	&	&Probability that the source is saturated\\
{\tt yHallMag}	&mag	&Point source Y mag\\
{\tt yHallMagErr}	&mag	&Error in point source Y mag\\
{\tt yPetroMag}	&mag	&Extended source Y mag (Petrosian)\\
{\tt yPetroMagErr}	&mag	&Error in extended source Y mag (Petrosian)\\
{\tt yPsfMag}	real	&mag	&Point source profile-fitted Y mag ({\bf not currently implemented})\\
{\tt yPsfMagErr}	&mag	&Error in point source profile-fitted Y mag ({\bf not currently implemented})\\
{\tt ySerMag2D}	&mag	&Extended source Y mag (profile-fitted) ({\bf not currently implemented})\\
{\tt ySerMag2DErr}	&mag	&Error in extended source Y mag (profile-fitted) ({\bf not currently implemented})\\
{\tt yAperMag3}	&mag	&Extended source Y mag (2.0 arcsec aperture diameter)\\
{\tt yAperMag3Err}	&mag	&Error in extended source Y mag (2.0 arcsec aperture diameter)\\
{\tt yAperMag4}	&mag	&Extended source Y mag (2.8 arcsec aperture diameter)\\
{\tt yAperMag4Err}	&mag	&Error in extended source Y mag (2.8 arcsec aperture diameter)\\
{\tt yAperMag6}	&mag	&Extended source Y mag (5.7 arcsec aperture diameter)\\
{\tt yAperMag6Err}	&mag	&Error in extended source Y mag (5.7 arcsec aperture diameter)\\
{\tt yGausig}		&pixels	&RMS of axes of ellipse fit in Y\\
{\tt yEll}		& 	&1-b/a, where a/b=semi-major/minor axes in Y\\
{\tt yPA}		&Degrees&	Ellipse fit celestial orientation in Y\\
{\tt yErrBits}	& 	&Processing warning/error bitwise flags in Y\\
{\tt yDeblend}	& 	&Flag indicating parent/child relation in Y\\
\hline
\end{tabular}
\caption{List of parameters in the LAS source table. Note that
attributes yHallMag to yEta repeat for each of j\_1, j\_2, h \& k --
these are omitted here for brevity.  Currently, all attributes relating to 
proper motion and profile fitting contain default values.}
\label{tab_source_table}
\end{table*}

\begin{table*}
\centering
\begin{tabular}{@{}lll@{}}
\hline
Name		&Unit	&Description	\\
\hline
{\tt yClass}		&	&Discrete image classification flag in Y\\
{\tt yClassStat}	& 	&N(0,1) stellarness-of-profile statistic in Y\\
{\tt yppErrBits}	& 	&Additional WFAU post-processing error bits in Y\\
{\tt ySeqNum}		&	&Running number of the Y detection\\
{\tt yObjID}		&	&UID of the Y detection\\
{\tt yXi} 	&arcsec	&Offset of Y detection from master position (+east/-west)\\
{\tt yEta}		&arcsec	&Offset of Y detection from master position (+north/-south)\\
\hline
\end{tabular}
\contcaption{List of parameters in the LAS source table. }
\end{table*}

\label{lastpage}

\end{document}